\documentclass[11pt]{article}
\usepackage{epsfig} 
\usepackage{amssymb} 
\usepackage{graphics}
\newcommand{\sect}[1]{ \section{#1} \setcounter{equation}{0} }

\newcommand{\third}{\mbox{\small{$\frac{1}{3}$}}} 
\newcommand{\threequarters}{\mbox{\small{$\frac{3}{4}$}}} 
\newcommand{\ninesixteenths}{\mbox{\small{$\frac{9}{16}$}}} 
\newcommand{\fourthirds}{\mbox{\small{$\frac{4}{3}$}}} 
\newcommand{\MSbar}{\overline{\mbox{MS}}} 
 
\newcommand{\Nc}{N_{\!c}}
\newcommand{\Nf}{N_{\!f}}

\begin{document}

\begin{centering}

\vfill

\hspace{9cm}
{\bf MaPhy-AvH/2019-09, LTH 1207}

\vspace{1.5cm}
{\LARGE {\bf On the self-consistency of off-shell Slavnov-Taylor identities in
QCD}}

\vspace{1.2cm}
\noindent
{\bf J.A. Gracey}

\vspace{0.2cm}
\noindent
{Theoretical Physics Division, Department of Mathematical Sciences, University 
of Liverpool, P.O. Box 147, Liverpool, L69 3BX, United Kingdom} 

\vspace{0.3cm}
{\bf H. Ki\ss{}ler \& D. Kreimer}

\vspace{0.2cm}
\noindent
{Department of Mathematics, Humboldt-Universit\"{a}t zu Berlin, Rudower
Chaussee 25, D-12489 Berlin, Germany}

\vspace{0.7cm}
\noindent
{August, 2019.} 

\end{centering}


\vspace{5cm} 
\noindent 
{\bf Abstract.} Using Hopf-algebraic structures as well as diagrammatic 
techniques for determining the Slavnov-Taylor identities for QCD we construct 
the relations for the triple and quartic gluon vertices at one loop. By making 
the longitudinal projection on an external gluon of a Green's function we show 
that the gluon self-energy of that leg is consistently replaced by a ghost 
self-energy. The resulting identities are then studied by evaluating all the 
graphs for an off-shell non-exceptional momentum configuration. In the case of 
the $3$-point function this is for the most general momentum case and for the 
$4$-point function we consider the fully symmetric point. 

\newpage 

\sect{Introduction}

One of the cornerstones of quantum field theory is the accommodation of
spin-$1$ gauge fields in the Lagrangian of a theory in such a way that the core
properties of the gauge field are retained. For instance, a gauge field
$A^a_\mu$ will describe a photon or gluon in the respective abelian or
non-abelian cases where the Lagrangian will be built from gauge invariant
operators of $A^a_\mu$. However, such an object has too many degrees of freedom
and to properly describe physical phenomena the gauge field needs to satisfy
constraints known as gauge conditions. The inclusion of such a condition in the
Lagrangian breaks gauge invariance which is one guiding principle behind
physical predictions. In the classical theory such gauge fixings do not lead
to insurmountable problems. For instance, performing computations in different
gauges will give the same physical outcome. In the quantum theory this is not
as straightforward since in covariant gauges, as an example, the choice of
gauge can change due to quantum corrections. Therefore it is not clear if
the remnant of the gauge symmetry, evident in the classical case, is also
preserved quantum mechanically. The development of the 
Becchi-Rouet-Stora-Tyutin (BRST) transformation put this problem on a firm 
footing in that a symmetry of the gauge fixed and hence gauge variant 
Lagrangian was constructed. One benefit was that it provided the machinery to 
confirm that the physical state space of the gauge field was positive definite 
ensuring that the Lagrangian satisfies unitarity. As equally as important as 
this is that the formalism reproduced the non-abelian extension of the 
Ward-Takahashi identities which are termed the Slavnov-Taylor identities, 
\cite{1,2}. Briefly these are relations between different $n$-point functions 
of the {\em quantum} theory and such relations have to hold in the bare and 
renormalized cases. In the latter situation this means that constraints on the 
renormalization constants implied by the identities have to be satisfied in 
each choice of renormalization scheme, \cite{1,2,3,4}. It is widely known that 
in Quantum Chromodynamics (QCD) that the coupling renormalization constant 
derived from one of the $3$- or $4$-point vertices in the modified minimal 
subtraction ($\MSbar$) scheme is automatically consistent with that derived 
from the remaining ones, \cite{3,5}. In other schemes this may not be the case.
So fixing the coupling constant renormalization from one vertex means that the 
structure of the other vertex functions is determined {\em using} the 
restrictions from the Slavnov-Taylor identities. This has been studied in depth
in QCD in a variety of early articles such as \cite{3,6,7,8,9,10,11}. More 
recently Slavnov-Taylor identities have been used to analyse the structure of 
$n$-point functions in order to probe the infrared dynamics of QCD. Various 
review articles, for instance, give a flavour of developments over the last 
decade, \cite{12,13,14,ver1}. More recently, progress in understanding the 
non-perturbative structure of the triple gluon vertex has been made through
Dyson-Schwinger, functional renormalization group and lattice methods,
\cite{ver2,ver3,ver4,ver5,ver6}. For example, a comprehensive study of the 
non-perturbative longitudinal part of the triple gluon vertex was provided in
\cite{ver7}. Although $3$-point QCD vertex studies have been the main focus, a 
similar level of non-perturbative analysis is becoming available for $4$-point 
vertices primarily through the Dyson-Schwinger technique,
\cite{ver8,ver9,ver10,ver11,ver12,ver13}. 

Diagrammatic techniques for the construction of QCD Slavnov-Taylor identities 
have been provided in \cite{1,3,15,16,17}. More recently these ideas have
been used in several articles, \cite{18,19}. For instance, in \cite{18} the
Hopf-algebraic structure of Slavnov-Taylor identities was examined with the
ghost sector being shown to have a connection with the Corolla polynomial. In
\cite{15,Kissler:2017phd} the diagrammatic approach was used to reorganize 
Feynman diagrams contributing to Dyson-Schwinger equations in Quantum 
Electrodynamics (QED). An important result was that the gauge parameter 
dependence of the electron propagator in a linear covariant gauge was 
reconstructed from a pure Feynman gauge analysis. The formalism was shown to be
correct to {\em four} loops. While the diagrammatic approach of 
\cite{1,3,15,16,17} is perhaps not a mainstream method since it does not use 
path integral methods or the technique of algebraic renormalization, \cite{20},
importantly it does preserve the distinction between the transverse and 
longitudinal components of the gauge field within Green's functions and allows 
one to follow their individual routes through a graph. One useful aspect of 
diagrammatic techniques is that Slavnov-Taylor identity-like relations between 
one-particle irreducible (1PI) Green's functions can be derived without 
explicitly studying connected Green's functions. Such 1PI Green's functions 
have been checked calculationally in several articles, \cite{10,11}. In 
\cite{10} the triple gluon vertex was studied in QCD in the linear covariant 
gauge and axial gauge at one loop. While the identity for the gluon $4$-point 
was discussed in \cite{3,10} it was not checked but one loop calculations were 
carried out in \cite{11}. In that latter article the quartic vertex was 
examined at the completely symmetric point which is a non-exceptional momentum 
configuration. Moreover the consequences of the Slavnov-Taylor identity for the
renormalization constants were studied in the Weinberg scheme, \cite{4}. More 
recently the analysis of \cite{11} was extended in \cite{21} where the full 
decomposition of the quartic vertex into all the Lorentz tensor and colour 
channels was given. Aside from a few minor typographical errors the expression 
found in \cite{11} for the Lorentz tensors purely corresponding to the quartic 
gluon Feynman rule was effectively correct. However one observation of 
\cite{11} was that the relations between renormalization constants were not 
satisfied as they ought to have been due to the Slavnov-Taylor implications. 
Choices of the gauge parameter were found to ameliorate the situation. 

Therefore to study the Slavnov-Taylor identities afresh we return to basics and
apply modern algebraic and diagrammatic methods to construct the identities of 
the various relevant $3$- and $4$-point functions. These will involve the 
triple gluon and ghost-gluon vertices and both the pure gluon and ghost-gluon 
$4$-point functions. The latter was studied in \cite{22} together with the 
other possible $4$-point functions of QCD at one loop at the symmetric point. 
While the $3$- and $4$-point ghost-gluon vertex functions have been studied in 
earlier work we have to carry out a new evaluation here. This is because in the
standard construction of the Slavnov-Taylor identity the vertex connecting to 
one of the external ghost fields is not the standard one derived using the 
Faddeev-Popov method, \cite{1,2,3}. Instead for that specific vertex the 
momentum appearing in the Feynman rule is stripped off to produce a vertex rule
with two Lorentz indices. Since the vertex function of this modified vertex is 
required for our computations we have to evaluate it for a completely off-shell
momentum configuration. We will also provide a general derivation of the
identities using Hopf-algebraic arguments based on \cite{anatomy} valid at all
orders in perturbation theory. In addition to this we will carry out explicit 
one loop calculations for each Slavnov-Taylor identity for an off-shell setup. 
In the case of the $3$-point identity this will be in the fully off-shell case 
while for the $4$-point one we will focus on the same fully symmetric point as 
\cite{11}. In both cases we will show that the identities are fully satisfied 
in all colour and Lorentz channels. While this appears to contradict the 
observation of \cite{11}, in our derivation using combinatorial Dyson-Schwinger 
equations, \cite{anatomy}, and the diagrammatic approach following 
\cite{1,3,15,16,17,18B}, additional graphs arise which appear to be absent or 
implicit in earlier work for 1PI Green's functions. Their presence is crucial 
to reconciling the identities. At this juncture our primary concern is to 
demonstrate the consistency of the 1PI Slavnov-Taylor identities. What the 
implications of the results are for other work still has to be followed 
through.

The paper is organized as follows. We devote Section $2$ to the description of
the Hopf-algebraic and diagrammatic constructions of the $3$- and $4$-point 
identities which we will study using explicit computations. The identity 
relating the triple gluon vertex to ghost-gluon $3$-point functions is studied 
in depth at one loop in the off-shell case in Section $3$. A similar analysis 
but for the $4$-point identity at the fully symmetric point is carried out in 
the next section with conclusions given in Section $5$. Several appendices are 
provided. These give the details of the projection matrices and tensor basis, 
the core colour group theory needed for the $4$-point function calculation with
the final appendix giving explicit expressions for the purely gluonic $3$- and 
$4$-point functions. 

\sect{Construction of identities.}

We devote this section to the derivation of the identities by exploiting
algebraic structures which originate due to combinatorial insertions of Green's
functions \cite{anatomy} and studying diagrammatic techniques following
\cite{1,3,15,16,17}. In particular we will offer two
different derivations of the desired identities. The main goal is to clarify 
how gauge symmetry can be expressed on the level of renormalized 1PI Green's 
functions and the resulting implications in the corresponding algebra of 
structure functions.

\subsection{Hopf-algebra derivation.}

Our first starting point is to consider the Dyson-Schwinger equations for the 
1PI Green's functions. The demand that QCD can be renormalized by the unique 
renormalization of a single coupling constant $g$ delivers a set of identities 
for the renormalization factor $Z_g$ of the coupling constant
\begin{equation}\label{ZSTcoupl}
Z_g ~=~ \frac{Z_{\Gamma^{ggg}}}{(Z_{\Gamma^{gg}})^{\frac{3}{2}}} ~=~
\frac{Z_{\Gamma^{g\bar{q}q}}}{Z_{\Gamma^{\bar{q}q}}\sqrt{Z_{\Gamma^{gg}}}} ~=~ 
\frac{Z_{\Gamma^{g\bar{c}c}}}{Z_{\Gamma^{\bar{c}c}}\sqrt{Z_{\Gamma^{gg}}}} ~=~ 
\frac{\sqrt{Z_{\Gamma^{gggg}}}}{Z_{\Gamma^{gg}}} ~.
\end{equation}
We note that at the outset our notation is that when we label the 
renormalization constants or Green's function to distinguish which $n$-point 
function they relate to we use the letters $g$, $c$ and $q$ to indicate gluons,
Faddeev-Popov ghosts and quarks respectively as well as the associated 
antiparticles in the latter two instances. So, for example, the label $gggg$
indicates the gluon $4$-point function or quartic gluon vertex function. A 
derivation of these identities (\ref{ZSTcoupl}) can be achieved using the 
locality of counterterms in a renormalizable field theory which implies that 
the all orders counterterms can be obtained from a solution of a fixed point 
equation in Hochschild cohomology \cite{anatomy} and we refer readers to that
article for background to the notation used for the derivation by this method.

As a consequence the relations (\ref{ZSTcoupl}) are obtained by applying the 
counterterm map $S_R^\Phi$ to combinatorial Green's functions. They themselves 
obey a similar formal identity
\begin{equation}\label{STcoupl}
\frac{\Gamma^{ggg}}{(\Gamma^{gg})^{\frac{3}{2}}} ~=~ 
\frac{\Gamma^{g\bar{q}q}}{\Gamma^{\bar{q}q}\sqrt{\Gamma^{gg}}} ~=~
\frac{\Gamma^{g\bar{c}c}}{\Gamma^{\bar{c}c}\sqrt{\Gamma^{gg}}} ~=~
\frac{\sqrt{\Gamma^{gggg}}}{\Gamma^{gg}} ~.
\end{equation}
Let us define 1PI combinatorial Green's functions 
\begin{equation}
\Gamma^r(g^2) ~:=~ t_r\pm \sum_{\mathbf{res}(\Gamma)=r}g^{2|\Gamma|}\frac{\Gamma}{|\mathrm{Aut}(\Gamma)|} ~=~
t_r\pm \sum_{k=1}^\infty g^{2k} B_+^{k;r}(\Gamma^r Q^{2k}),
\end{equation}
where $r\in\mathcal{R}$ specifies the 1PI amplitude under consideration. For us
it suffices to consider 
\[
\mathcal{R} ~=~ \{\bar{c}c,gg,ggg,gggg,g\bar{c}c,\bar{c}cgg\}
\]
the inverse ghost and gluon propagators, the $3$- and $4$-gluon vertex 
functions, and the coupling of one or two gluons with a ghost pair.
We take $t_r$~$=$~$\Gamma^r_{(0)}$ to be the tree-level contribution for such 
an amplitude. It can vanish as it does in $t_{\bar{c}cgg}$~$=$~$0$, as there is
no quartic gluon-ghost interaction in the linear covariant gauge fixed
Lagrangian. Indeed there is no need for such a term as it does not have to be 
renormalized as an overall convergent contribution.
Furthermore, $\mathbf{res}(\Gamma)$ is obtained by shrinking internal edges in
1PI graphs to zero length.
The fact that the $B_+^{k;r}$ act as Hochschild $1$-cocycles ensures the 
desired renormalization by local counterterms. For the maps $B_+^{k;r}$ to be 
indeed closed in Hochschild cohomology the identities (\ref{STcoupl}) are 
necessary and sufficient \cite{anatomy}.
Note that the 1PI $2$-point functions are inverse propagators, in particular
\begin{equation}\label{self}
\Gamma^{gg}(g^2) ~=~ 1-\tilde{\Gamma}^{gg}(g^2) ~~,~~ 
\Gamma^{\bar{c}c}(g^2) ~=~ 1-\tilde{\Gamma}^{\bar{c}c}(g^2)
\end{equation}
where $\tilde{\Gamma}$ indicates self-energies.

{\begin{figure}[hb]
\begin{center}
\includegraphics[width=14.0cm,height=16.0cm]{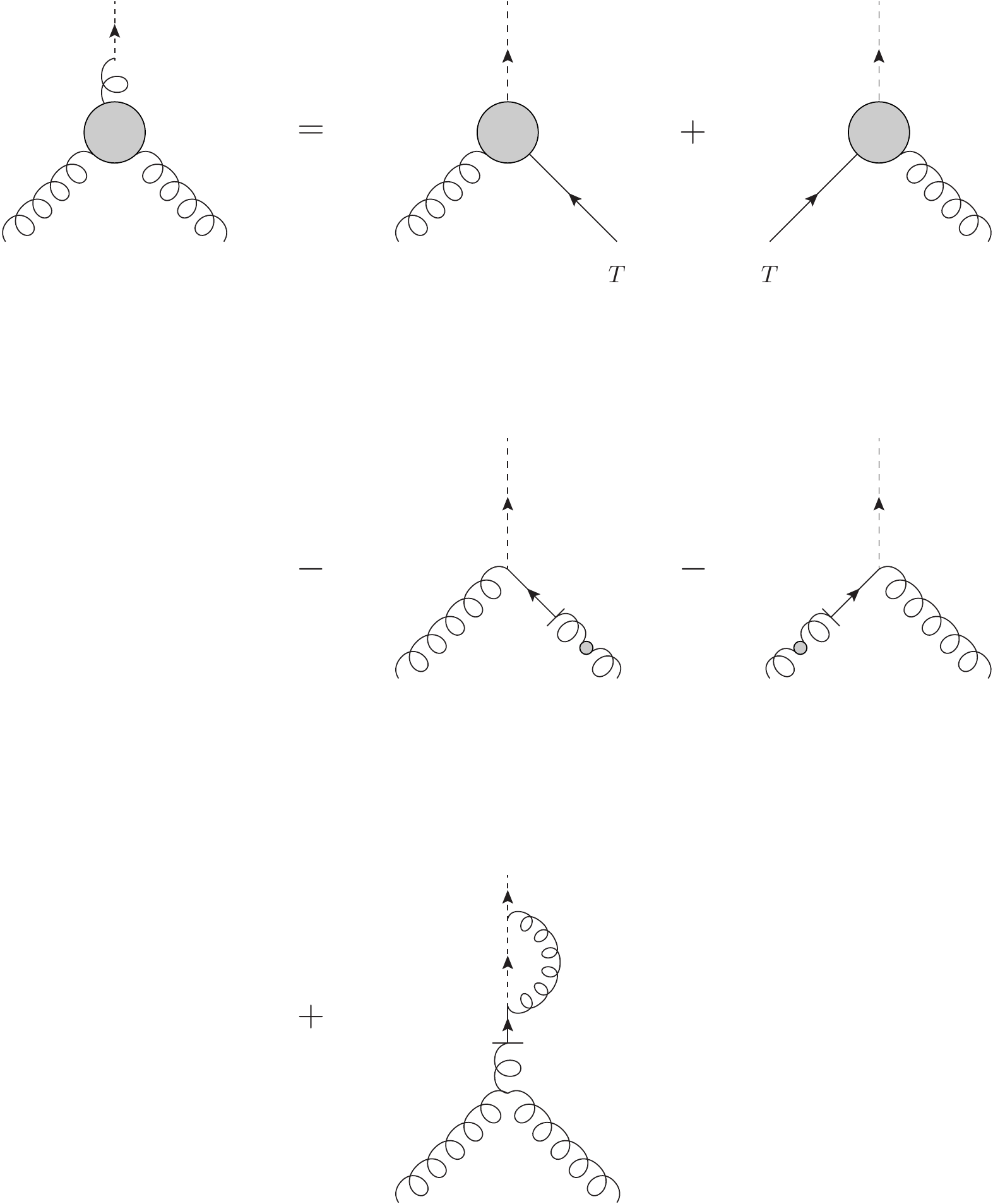}
\end{center}
\caption{Slavnov-Taylor identity for $3$-point function.}
\label{fig3pt}
\end{figure}}

Two further remarks are in order. The product of combinatorial Green's 
functions always implies a product as connected diagrams with a sum over all 
orientations understood. This is required by Hochschild cohomology where 
products of combinatorial Green's functions appear in arguments of the 
$1$-cocycles $B_+^{k;r}$, and closedness can only be achieved when the correct 
sum over all orientations is taken into account \cite{anatomy,KvS}.
As an example, 
\[
\frac{\Gamma^{gggg}}{\Gamma^{ggg}} ~=~ 
\frac{\Gamma^{ggg}}{\Gamma^{gg}} ~~ \Leftrightarrow ~~ 
\Gamma^{gggg} ~=~ \Gamma^{ggg}\cdot \frac{1}{\Gamma^{gg}}\cdot \Gamma^{ggg}
\]
where the propagator $\frac{1}{\Gamma^{gg}}$ is sandwiched between two 
$3$-gluon vertex functions in the three $s$, $t$ and $u$ topologies, with
respect to the Mandelstam variables, as indicated by the $\cdot$ notation.
Also understood is a form factor decomposition whenever appropriate. To allow 
for a projection onto chosen form factors, we extend the notion of a graph 
$\Gamma$ to a pair $(\Gamma,\sigma)$ where $\sigma\in \{\varsigma\}$, with 
$\{\varsigma\}$ a complete basis for the form factor decomposition of the 
evaluation of $\Gamma$ under renormalized Feynman rules.
The graph insertion is stable under projection onto a chosen form factor 
$\sigma$ upon summing over the complete basis for the inserted graphs $\gamma$
\begin{equation}
\sum_{\tilde{\sigma}}(\bar{\Gamma},\sigma)*(\gamma,\tilde{\sigma}) ~=~
\sum_\Gamma 
\frac{n(\bar{\Gamma},\gamma,\Gamma)}{|\gamma|_\wedge}(\Gamma,\sigma)
\end{equation}
in the notation of \cite{anatomy}. This ensures that projection onto a desired 
form factor commutes with replacing an edge or vertex by a full propagator or 
vertex Green's function.
 
The identities (\ref{STcoupl}) above constitute several co-ideals in 
accordance with Hochschild cohomology
\begin{equation}
\frac{\Gamma^{ggg}}{\Gamma^{gg}} ~=~ 
\frac{\Gamma^{g\bar{c}c}}{\Gamma^{\bar{c}c}} ~=~ 
\frac{\Gamma^{g\bar{q}q}}{\Gamma^{\bar{q}q}} ~=~
\frac{\Gamma^{gggg}}{\Gamma^{ggg}} ~.
\end{equation}
Of particular interest is the equation constituted by the first equality
\begin{equation}\label{coidealt}
\Gamma^{\bar{c}c}\cdot \Gamma^{ggg} ~=~ \Gamma^{g\bar{c}c}\cdot \Gamma^{gg}
\end{equation}
which implies
\begin{equation}\label{coidealf}
\Gamma^{\bar{c}c}\cdot \Gamma^{gggg} ~=~ 
\underbrace{\Gamma^{\bar{c}cgg} ~+~ 
\Gamma^{g\bar{c}c}\cdot\Gamma^{ggg}}_{=\Gamma^{\bar{c}cgg}_c}
\end{equation}
where $\Gamma^{\bar{c}cgg}_c$ is a connected combinatorial Green's function 
from
\begin{equation}\label{eqggcc}
\Gamma^{\bar{c}c}\cdot \Gamma^{ggg}\cdot 
\frac{1}{\Gamma^{gg}}\cdot \Gamma^{ggg}~=~ \Gamma^{g\bar{c}c}\cdot
\frac{\Gamma^{gg}}{\Gamma^{gg}}\cdot \Gamma^{ggg} ~=~ 
\Gamma^{g\bar{c}c}\cdot\Gamma^{ggg}
\end{equation}
using (\ref{coidealt}). The Green's function $\Gamma^{\bar{c}cgg}_c$ as a 
1PI Green's function contributing to the same amplitude as the connected 
Green's function is (\ref{eqggcc}). It hence must be included.

Upon using (\ref{self}) and expanding in $g^2$, (\ref{coidealt}) formally
becomes
\begin{equation}
\Gamma^{ggg}_{(1)} ~=~ \tilde{\Gamma}^{\bar{c}c}_{(1)}\cdot 
\Gamma^{ggg}_{(0)} ~+~ \Gamma^{g\bar{c}c}_{(1)}\cdot 
\underbrace{P}_{\equiv \Gamma^{gg}_{(0)}} ~-~ 
\Gamma^{g\bar{c}c}_{(0)}\cdot \tilde{\Gamma}^{gg}_{(1)}
\label{hoch3pt}
\end{equation}
where
\begin{equation}
P_{\mu\nu}(p) ~=~ \eta_{\mu\nu} ~-~ \frac{p_\mu p_\nu}{p^2}
\end{equation}
is the transverse projector and this is the first desired identity. It is 
illustrated in Figure \ref{fig3pt} and given in more explicit detail in 
(\ref{3ptid}), where a projection onto a longitudinal component 
$p_\sigma$ for a fixed chosen external leg is automatic on both 
sides above. We note that in Figure \ref{fig3pt} a blob at a vertex represents 
all 1PI one loop contributions and a gluon leg with a blob indicates the one 
loop corrections to the $2$-point function. Similarly, using (\ref{self}) again
and expanding in $g^2$, (\ref{coidealf}) becomes
\begin{equation}
\Gamma^{gggg}_{(1)} ~=~ 
\tilde{\Gamma}^{\bar{c}c}_{(1)}\cdot \Gamma^{gggg}_{(0)} ~+~ 
\Gamma^{\bar{c}cgg}_{(1)} 
~+~ \Gamma^{g\bar{c}c}_{(1)}\cdot P\cdot \Gamma^{ggg}_{(0)} 
~+~ \Gamma^{g\bar{c}c}_{(0)}\cdot P\cdot \Gamma^{ggg}_{(1)}
\label{hoch4pt}
\end{equation}
which is the second desired identity. A sum over orientations is understood in 
both equations so that they are indeed in complete agreement with Figures 
\ref{fig3pt} and \ref{fig4pt}. Again, a projection onto the longitudinal 
component $p_\sigma$ on a chosen external leg is automatic. We have illustrated
our notation for the various edges in Figure \ref{fig2pt} together with their 
various Feynman rules. In each rule the through momentum is $p$ and we have 
omitted the unit matrix in the colour indices.

\vspace{0.3cm}
{\begin{figure}
\begin{center}
\includegraphics[width=13.0cm,height=17.5cm]{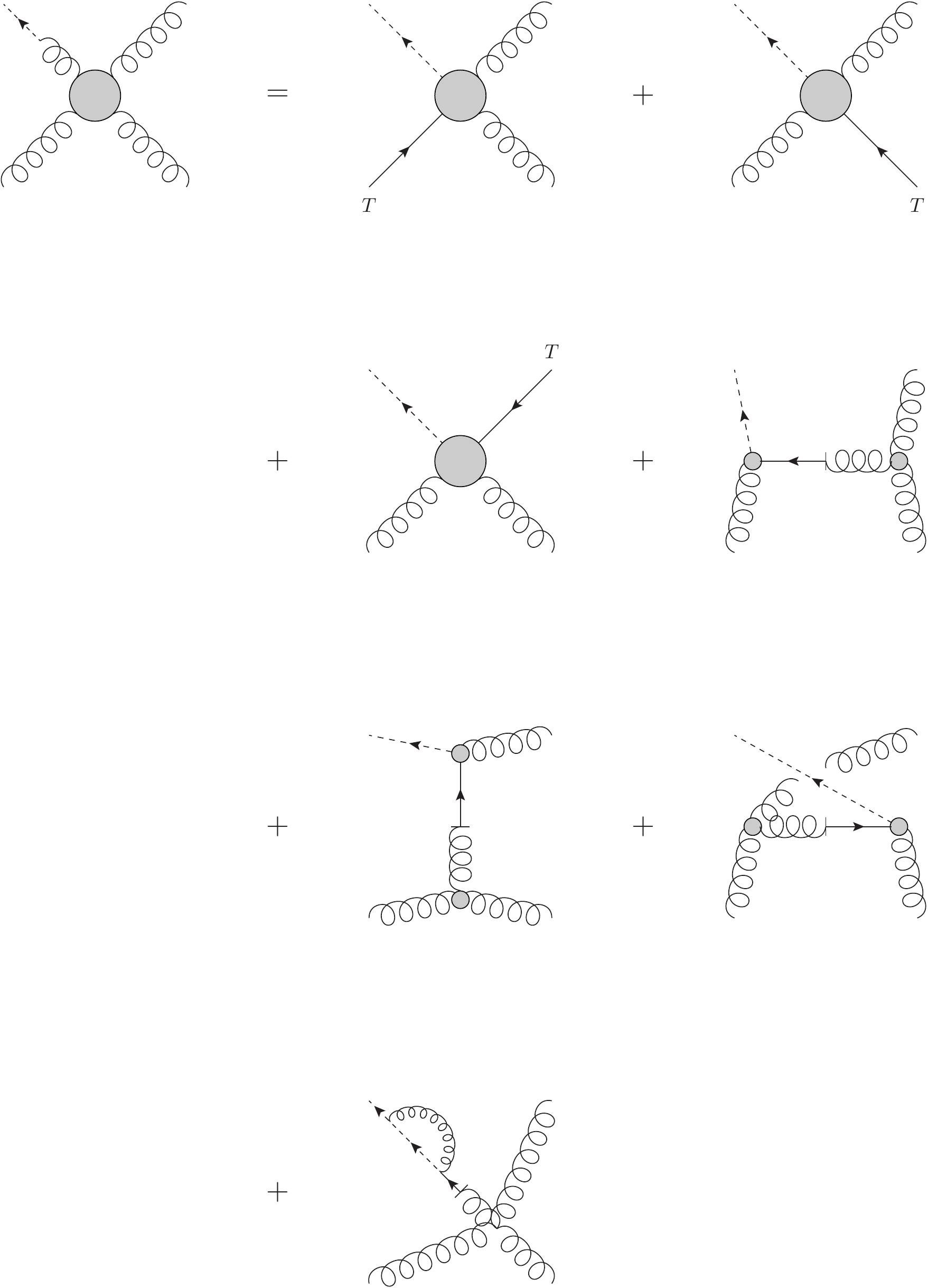}
\end{center}
\caption{Slavnov-Taylor identity for $4$-point function.}
\label{fig4pt}
\end{figure}}

The same result can be obtained by studying the contraction of a connected 
Green's function with $n$ external gluons at a fixed external gluon leg $i$ 
with its momentum $p_i$. The Slavnov--Taylor identity is 
\begin{equation}
p_i\cdot G^{n}(p_1,\ldots,p_n) ~=~ 0 ~.
\end{equation}
From \cite{18} we know that all graphs contributing to such a connected 
amplitude can be obtained by applying the Corolla polynomial \cite{18A} to a
corresponding sum of $3$-regular scalar graphs. Underlying this is a bi-complex
in graph and cycle homology studied in \cite{18} which puts the approach of 
\cite{15,16,17} on a firm mathematical footing. A careful rederivation of the 
Slavnov-Taylor identities using this approach is given in \cite{18B}. In
particular see Lemma 5.9 there which allows one to follow the resulting 
propagation of the corresponding longitudinal momenta through the graphs. If we
dress an external gluon leg of a 1PI vertex function in QCD by a gluon 
self-energy and contract with the gluon momentum, properties of the Corolla 
polynomial, discussed in Sections 6.1 and 6.9 of \cite{18}, ensure that this 
results in a 1PI vertex function where that external leg is longitudinal and 
dressed by a ghost self-energy. Indeed the Feynman graphs for the latter pair 
off with the sum of all paths through a gluon self-energy. This again leads to 
the desired identities.

\subsection{Diagrammatic derivation.}

Our first derivation of the Slavnov-Taylor identities clearly shows how the 
underlying algebra implies restrictions on the renormalization of Green's 
functions to all loop orders. In order to provide an illustration of the 
derived identities as well as a practical alternative to complement the general 
argument, and which in fact was the original way we discovered the relations
(\ref{hoch3pt}) and (\ref{hoch4pt}), we now examine one loop diagrams and 
discuss the diagrammatic approach following \cite{1,3,15,16,17}. 

{\begin{figure}[hb]
\begin{center}
\includegraphics[width=7.0cm,height=5.0cm]{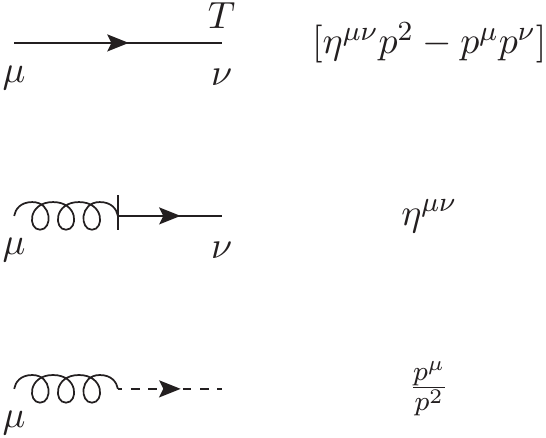}
\end{center}
\caption{Notation for graph representation of Slavnov-Taylor identities.}
\label{fig2pt}
\end{figure}}

For the concrete evaluation of the diagrams below, we are mainly concerned with
two basic identities that either link the $3$-gluon 1PI Green's function 
$\Gamma^{ggg}$ or the $4$-gluon 1PI Green's functions $\Gamma^{gggg}$ to a 
linear combination of certain connected Green's functions. The procedure for 
achieving this for one loop diagrams begins with contracting one of the 
external gluon legs with its in-going momentum. We will refer to this here as 
the longitudinal contraction which corresponds to the final rule in Figure
\ref{fig2pt}. The next stage is to examine the effect this contraction has on 
each individual graphs of the Green's function. As each contributing diagram is 
1PI and has amputated external propagators, the longitudinal contraction of the
external gluon leg is always incident to a vertex. At this stage we need to
distinguish the effect the contraction has on each of the possible vertices of 
the linear covariant gauge fixed QCD Lagrangian we focus on in this article.

First, if the incident vertex is of quark-gluon type then the diagrammatic 
cancellations which are applied are very similar to the abelian case that was 
discussed in \cite{Kissler:2017phd,Kissler:2018lnn}. However, due to the 
presence of the non-abelian group generator in the quark-gluon vertex we need 
to consider gauge invariant sets of graphs as discussed in 
\cite{Cvitanovic:1980bu}. This includes diagrams which are no longer 1PI. They 
will feature connected diagrams with a bridge corresponding to the contracted 
propagator of the final two terms of (\ref{hoch4pt}) or the third and fourth 
graphs on the right hand side of Figure \ref{fig3pt} for instance. Next if the 
incident vertex is a triple gluon vertex then it is straightforward to see that 
using the rules given in \cite{15,16,17,18B} the contracted vertex
can be rewritten in terms of an auxiliary ghost vertex allowing us to recognize
that the longitudinal gluon momentum propagates to the next adjacent vertex or 
contracts the propagator adjacent to both of these vertices. Repeating the
application of the rules means that the longitudinal gluon momentum propagates 
through a diagram until it hits a vertex which is not a triple gluon vertex or 
eventually reaches an external leg. The former diagram can be shown to cancel 
1PI diagrams with a contracted propagator that emerges from longitudinal 
contractions of a $4$-gluon vertex which is discussed below. The latter 
diagrams constitute a new type of Green's function where the external leg that
has been reached equals an in-going ghost leg of an incident vertex which has
a purely transverse component as derived in \cite{1}. The corresponding fixed 
longitudinal gluon leg is then identified with an out-going ghost leg. An
illustration of this is given, for example, in the first and second Green's 
functions on the right hand side of Figure \ref{fig3pt}. In terms of 
(\ref{hoch3pt}) and (\ref{hoch4pt}) this is consistent with the transverse
projection there.

Next for the case when the incident vertex involves a ghost vertex then it is 
possible to show that a certain linear combination of diagrams with internal 
ghost loops together with diagrams which have an external ghost line that 
originated from the longitudinal contraction of the fixed gluon vanishes. The 
essence of this cancellation of the ghost loops and longitudinal lines resides 
in the Jacobi identity for the structure constants. The final situation we have
to consider is that where the incident vertex involves the quartic gluon 
vertex. Then the $4$-valent vertex gets replaced by two $3$-valent vertices 
that are connected by a contracted propagator edge, \cite{16,17,18}. To be more
precise, all $(2|2)$ partitions of the four edges of the $4$-gluon vertex to 
the two emerging vertices need to be considered. From graph homology these 
partitions are well known and termed IHX terms and more familiarly correspond 
to the $s$, $t$ and $u$ channels in the Mandelstam variable notation. For full 
details on the quartic gluon vertex identity, we refer the reader to \cite{18} 
which also includes a detailed account of graph homology in QCD that underlies 
these identities. As discussed in the case of the gluon $3$-point case, the 
contracted propagator edge that connected the new $3$-valent vertices can be 
arranged to cancel contributions from other diagrams as long it is not a 
bridge. By contrast if it is a bridge then the contracted propagator edge 
contributes a new type of connected Green's function to the identity. Examples 
of this are evident in either the third, fourth or fifth graph on the right 
hand side of Figure \ref{fig3pt} or the third, fourth or fifth graphs on the 
right hand side of Figure \ref{fig4pt}. This explains the extra diagram of the
gluon $3$-point function $\Gamma^{ggg}$ in Figure \ref{fig3pt}. A separate case 
that we need to consider is the circumstance that occurs when an edge that is 
incident to a $4$-gluon vertex is contracted. This can be resolved since there 
is a simple cancellation rule that follows from the Jacobi identity as 
demonstrated in \cite{18,18B}. As a result the sum over all ways to 
contract one of the edges of the $4$-gluon vertex vanishes. Therefore, all 1PI 
diagrams with a contracted edge incident to a $4$-gluon vertex vanish on the 
right hand side of the Slavnov-Taylor identity in Figure \ref{fig4pt}. However,
a single diagram remains since the contracted edge is a bridge. Therefore this 
explains the appearance of the last diagram of Figure \ref{fig4pt} in full 
accord with a similar origin in the previous Hochschild construction.

\sect{Triple gluon vertex.}

In deriving both Slavnov-Taylor identities by algebraic and diagrammatic 
methods we have arrived at the same relations. However in comparing our 
expressions with relations between similar Green's functions provided in, say, 
\cite{10} we note that for both cases we have an additional graph which is not 
1PI but involves self-energy corrections to the ghost on the same external leg 
corresponding to the longitudinal projection. In other words this Faddeev-Popov
ghost is intimately tied to the longitudinal gluon. This is already well known 
in the $2$-point context since the ghost is necessary to cancel unphysical 
degrees of freedom in the longitudinal sector and ensure the gluon $2$-point 
function is transverse. Therefore we now turn to explicit computations to 
demonstrate how important this extra graph is to ensuring our relations are 
consistent. We will carry this out for the cases where the momentum of none of 
the external legs to set to zero and focus on general non-exceptional momentum 
configurations. In the case of the $3$-point relation we will do so for the 
completely off-shell configuration. This will build on earlier work of 
\cite{7,8,23,24,25,26} where in the latter the two loop off-shell QCD $3$-point
vertex functions were computed. However it is not possible to immediately lift 
even the one loop vertex functions from \cite{26} to effect an immediate check 
on our $3$-point identity. This is because like \cite{1,2,10} the ghost-gluon 
vertex function of the identity is a modification of the corresponding vertex 
function of the Lagrangian. We note that the Feynman rule for the canonical 
ghost-gluon vertex for the linear covariant gauge we use involves the momentum 
of one of the ghost fields. The associated connected vertex function would then
be denoted by $\Gamma^{g\bar{c}c}_\mu(p,q,r)$ where the Lorentz index matches 
that of the gluon field and $p$, $q$ and $r$ are the external momenta. However 
as we have noted in the derivation of the identities an adjusted ghost-gluon 
vertex plays the major role. It is related to the canonical ghost-gluon vertex 
Feynman rule but with the external ghost momentum dropped. In general this 
vertex function is denoted by $\Gamma^{g\bar{c}c}_{\mu\nu}(p,q,r)$ and is 
graphically defined in Figure \ref{figaccpr}. The second Lorentz index is the 
place where the external ghost momentum would be attached to produce the 
canonical vertex present in the Lagrangian. Therefore the Feynman rule for 
$\Gamma^{g\bar{c}c}_{\mu\nu}(p,q,r)$ is proportional to $\eta_{\mu\nu}$ with 
the colour group and other factors remaining unchanged. In Figure 
\ref{figaccpr} we have included two ghost-gluon Green's functions labelled 
separately by $A$ and $B$. This is because both orientations appear in the 
identity and we need to be careful in computing both off-shell. The other 
Green's function of Figure \ref{figaccpr} defines the triple gluon vertex 
function of the identity which was computed in \cite{23,26}. We refer the 
reader to \cite{26} for the result with the same conventions used here. 

\vspace{0.3cm}
{\begin{figure}[ht]
\begin{center}
\includegraphics[width=16.0cm,height=5.5cm]{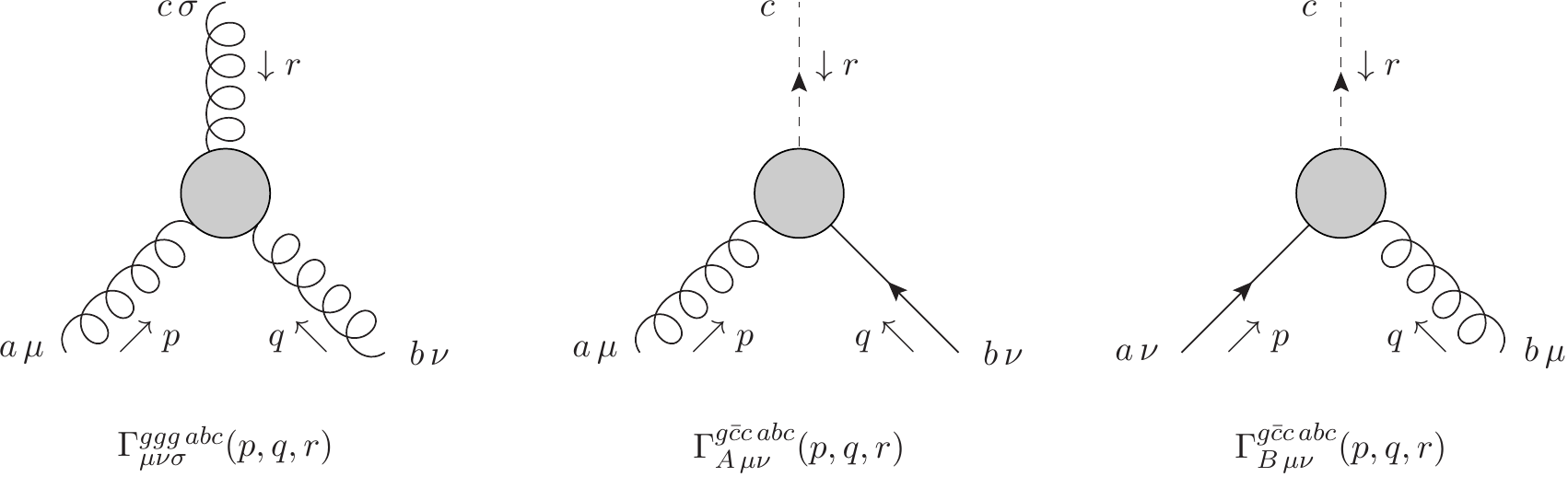}
\end{center}
\caption{Basic one particle irreducible Green's functions for $3$-point 
identity.}
\label{figaccpr}
\end{figure}}

While the ghost-gluon vertex function itself was also computed in \cite{23,26} 
we will need the off-shell result for $\Gamma^{g\bar{c}c}_{\mu\nu}(p,q,r)$ for 
both orientations. We briefly summarize the computation of \cite{26} here first 
noting that we used the same projection principle to decompose the vertex
function into a basis of Lorentz tensors. Using $r$ in each instance of a
$3$-point function as the dependent external momentum then there are now five 
possible tensors for both versions of $\Gamma^{g\bar{c}c}_{\mu\nu}(p,q,r)$,
rather than two for the conventional vertex function, which are 
\begin{eqnarray}
{\cal P}^{g\bar{c}c}_{(1) \mu\nu}(p,q) &=& \eta_{\mu\nu} ~~~,~~~
{\cal P}^{g\bar{c}c}_{(2) \mu\nu}(p,q) ~=~ \frac{p_\mu p_\nu}{\mu^2} ~~~,~~~
{\cal P}^{g\bar{c}c}_{(3) \mu\nu}(p,q) ~=~ \frac{p_\mu q_\nu}{\mu^2}  \nonumber \\
{\cal P}^{g\bar{c}c}_{(4) \mu\nu}(p,q) &=& \frac{q_\mu q_\nu}{\mu^2} ~~~,~~~
{\cal P}^{g\bar{c}c}_{(5) \mu\nu}(p,q) ~=~ \frac{q_\mu q_\nu}{\mu^2}
\end{eqnarray}
for the orientations of Figure \ref{figaccpr}. The kinematic variables for our
off-shell analysis are the same as \cite{27} and we note that
\begin{equation}
x ~=~ \frac{p^2}{r^2} ~~~,~~~ y ~=~ \frac{q^2}{r^2} ~~~,~~~ r^2 ~=~ -~ \mu^2
\label{3ptvar}
\end{equation}
when we consider $3$-point Green's functions. The associated Gram determinant
is, \cite{27},
\begin{equation}
\Delta_G(x,y) ~=~ x^2 ~-~ 2 x y ~+~ y^2 ~-~ 2 x ~-~ 2 y ~+~ 1 ~.
\end{equation}
With this basis then for route $A$ we have
\begin{equation} 
\left\langle A^a_\mu(p) c^b(q) \bar{c}_\nu^c(r) \right\rangle ~=~ 
f^{abc} \Gamma^{g\bar{c}c}_{A\,\mu\nu}(p,q,r) ~=~
f^{abc} \sum_{k=1}^{5} {\cal P}^{g\bar{c}c}_{A(k) \mu\nu}(p,q) \,
\Sigma^{g\bar{c}c}_{A(k)}(p,q)
\label{accpradecomp}
\end{equation} 
with
\begin{equation}
p ~+~ q ~+~ r ~=~ 0 
\end{equation}
and we have factored off the common colour group structure constants $f^{abc}$
which play a passive role at one loop. The Lorentz index on the field 
$\bar{c}^a_\mu$ indicates the removal of the external momentum from the 
associated ghost-gluon vertex. We also define
\begin{equation} 
\left\langle A^a_\mu(p) A^b_\nu(q) A^c_\sigma(r) \right\rangle ~=~ 
f^{abc} \Gamma^{ggg}_{\mu\nu\sigma}(p,q,r) ~.
\end{equation} 
The associated scalar amplitudes are $\Sigma^{g\bar{c}c}_{A(k)}(p,q)$ and are 
deduced by multiplying $\Gamma^{g\bar{c}c}_{A\,\mu\nu}(p,q,r)$ by the 
projection matrix ${\cal M}^{g\bar{c}c}_{kl}$ which is the inverse of the 
matrix
\begin{equation}
{\cal N}^{g\bar{c}c}_{kl} ~=~ {\cal P}^{g\bar{c}c}_{(k) \, \mu\nu}(p,q) 
{\cal P}^{{g\bar{c}c} \, \mu\nu}_{(l)}(p,q) 
\end{equation}
where $1$~$\leq$~$k$~$\leq$~$5$. The explicit expressions for the elements of 
${\cal M}^{g\bar{c}c}_{kl}$ are provided in Appendix A as they are complicated 
functions of the variables $x$ and $y$ as well as the spacetime dimension $d$. 
We note that throughout we use dimensional regularization to carry out all our
loop calculations in $d$~$=$~$4$~$-$~$2\epsilon$ dimensions. So we have to
carry out the projection in $d$-dimensions. To evaluate the vertex functions
to one loop we use the same method outlined in detail in \cite{26} and refer
the reader to that for more technical details. Though to summarize we note
that the integration by parts algorithm devised by Laporta, \cite{28}, is the
main tool and we used the {\sc Reduze} implementation, \cite{29,30}. The 
underlying master integrals are imported from \cite{31,32} to complete the
decomposition into scalar amplitudes. All our calculations for both $3$- and 
$4$-point functions are carried out automatically with the symbolic
manipulation language {\sc Form}, \cite{33,34}, used to handle the algebraic 
manipulations after the contributing Feynman graphs are generated using the 
{\sc Qgraf} package, \cite{35}.

To get a flavour of the consequences of our computations we record the 
expressions for both orientations of $\Gamma^{g\bar{c}c}_{\mu\nu}(p,q,r)$ in 
Figure \ref{figaccpr}. However given that the completely off-shell results 
involve polylogarithms we provide the versions at the completely symmetric 
point defined by $x$~$=$~$y$~$=$~$1$ for illustration. We have 
\begin{eqnarray}
-~ i \left. \Gamma^{g\bar{c}c}_{A\,\mu\nu}(p,q,r) \right|_{x=y=1} &=&
-~ \eta_{\mu\nu} g \nonumber \\
&& +~ 
\left[
\left[
\left[ 
\frac{\xi}{2} 
- \frac{1}{2}
\right]
\frac{1}{\epsilon} 
- 1
- \frac{\pi^2}{9}
+ \frac{\xi}{2} 
+ \frac{7\pi^2}{54} \xi
+ \frac{\xi^2}{8}
\right. \right. \nonumber \\
&& \left. \left. ~~~~~~~~
+ \frac{1}{6} \psi^\prime(\third)
- \frac{7}{36} \psi^\prime(\third) \xi
\right]
\eta_{\mu\nu} 
\right. \nonumber \\
&& \left. ~~~~~
+ 
\left[ 
\frac{4 \pi^2}{27}
- \frac{\xi}{6}
+ \frac{\pi^2}{27} \xi 
- \frac{\pi^2}{54} \xi^2
- \frac{2}{9} \psi^\prime(\third)
- \frac{\xi}{18} \psi^\prime(\third) 
\right. \right. \nonumber \\
&& \left. \left. ~~~~~~~~
+ \frac{\xi^2}{36} \psi^\prime(\third)
\right]
\frac{p_\mu p_\nu}{\mu^2} 
+
\left[ 
\frac{\xi}{6} 
+ \frac{\pi^2}{27} \xi
- \frac{\xi}{18} \psi^\prime(\third) 
\right]
\frac{p_\mu q_\nu}{\mu^2} 
\right. \nonumber \\
&& \left. ~~~~~
+ 
\left[ 
\frac{4 \pi^2}{27}
+ \frac{\xi}{6}
- \frac{\pi^2}{27} \xi
- \frac{\xi^2}{4}
- \frac{\pi^2}{27} \xi^2
- \frac{2}{9} \psi^\prime(\third)
+ \frac{\xi}{18} \psi^\prime(\third) 
\right. \right. \nonumber \\
&& \left. \left. ~~~~~~~~
+ \frac{1}{18} \psi^\prime(\third) \xi^2
\right]
\frac{p_\nu q_\mu}{\mu^2} 
\right. \nonumber \\
&& \left. ~~~~~
+ 
\left[ 
\frac{1}{3} \xi
- \frac{2}{27} \pi^2
+ \frac{1}{27} \pi^2 \xi
+ \frac{1}{9} \psi^\prime(\third)
- \frac{1}{18} \psi^\prime(\third) \xi
\right]
\frac{q_\mu q_\nu}{\mu^2} 
\right] C_A g^3 \nonumber \\
&& +~ O(g^5) 
\end{eqnarray}
and
\begin{eqnarray}
-~ i \left. \Gamma^{g\bar{c}c}_{B\,\mu\nu}(p,q,r) \right|_{x=y=1} &=&
\eta_{\mu\nu} g \nonumber \\
&&
+ \left[
\left[
\left[
\frac{1}{2}
- \frac{\xi}{2} 
\right] \frac{1}{\epsilon}
+ 1
+ \frac{\pi^2}{9}
- \frac{\xi}{2}
- \frac{7\xi}{54} \pi^2
- \frac{\xi^2}{8}
\right. \right. \nonumber \\
&& \left. \left. ~~~~~~~~
- \frac{1}{6} \psi^\prime(\third)
+ \frac{7 \xi}{36} \psi^\prime(\third)
\right]
\eta_{\mu\nu}
\right. \nonumber \\
&& \left. ~~~~~
+
\left[
\frac{2 \pi^2}{27}
- \frac{1}{3} \xi
- \frac{1}{27} \pi^2 \xi
- \frac{1}{9} \psi^\prime(\third)
+ \frac{1}{18} \psi^\prime(\third) \xi
\right]
\frac{p_\mu p_\nu}{\mu^2}
\right. \nonumber \\
&& \left. ~~~~~
+ 
\left[
- \frac{4 \pi^2}{27}
- \frac{\xi}{6}
+ \frac{\pi^2}{27} \xi
+ \frac{1}{4} \xi^2
+ \frac{\pi^2}{27} \xi^2
+ \frac{2}{9} \psi^\prime(\third)
- \frac{\xi}{18} \psi^\prime(\third)
\right. \right. \nonumber \\
&& \left. \left. ~~~~~~~~
- \frac{1}{18} \psi^\prime(\third) \xi^2
\right]
\frac{p_\mu q_\nu}{\mu^2}
+ 
\left[
- \frac{1}{6} \xi
- \frac{1}{27} \pi^2 \xi
+ \frac{1}{18} \psi^\prime(\third) \xi
\right]
\frac{p_\nu q_\mu}{\mu^2}
\right. \nonumber \\
&& \left. ~~~~~
+ 
\left[
- \frac{4}{27} \pi^2
+ \frac{1}{6} \xi
- \frac{1}{27} \pi^2 \xi 
+ \frac{1}{54} \pi^2 \xi^2
+ \frac{2}{9} \psi^\prime(\third)
+ \frac{1}{18} \psi^\prime(\third) \xi
\right. \right. \nonumber \\
&& \left. \left. ~~~~~~~~
- \frac{1}{36} \psi^\prime(\third) \xi^2
\right]
\frac{q_\mu q_\nu}{\mu^2}
\right] C_A g^3 ~+~ O(g^5)
\end{eqnarray}
where $\alpha$~$=$~$1$~$-$~$\xi$ is the gauge parameter with $\alpha$~$=$~$0$
corresponding to the Landau gauge, $\psi(z)$ is the derivative of the logarithm
of Euler $\Gamma$-function\footnote{The presence of $\psi(\third)$ is not
unrelated to the cyclotomic polynomials of \cite{cyclo}.}, $C_A$ is the usual 
colour group Casimir and $g$ is the gauge coupling constant. We do not carry 
out any renormalization at any instance since the identities hold in the bare 
case. So the parameters $\xi$, $\alpha$ and $g$ are bare. We have provided the 
expressions for these off-shell vertex functions in the attached data file. 

At this point it is worth recording the $3$-point Slavnov-Taylor identity for
the triple gluon vertex, illustrated in Figure \ref{fig3pt}, as an equation 
with the Lorentz indices explicit, now that we have introduced the modified 
ghost-gluon vertex $\Gamma^{g\bar{c}c}_{\mu\nu}(p,q,r)$. We have 
\begin{eqnarray}
r^\sigma \Gamma^{ggg}_{\mu\nu\sigma}(p,q,r) &=&
\Gamma^{g\bar{c}c}_{A\,\mu\rho}(p,q,r) P^\rho_{~\nu}(q) q^2 ~+~
\Gamma^{g\bar{c}c}_{B\,\nu\rho}(p,q,r) P^\rho_{~\mu}(p) p^2 \nonumber \\
&& -~ \Gamma^{g\bar{c}c}_{(0)\,A\,\mu\rho}(p,q,r) P^\rho_{~\nu}(p) 
\Gamma^{gg}_{(1)}(p) p^2 x^{-\epsilon} ~-~ 
\Gamma^{g\bar{c}c}_{(0)\,B\,\nu\rho}(p,q,r) P^\rho_{~\mu}(q) 
\Gamma^{gg}_{(1)}(q) q^2 y^{-\epsilon} \nonumber \\
&& +~ r^\sigma \Gamma^{ggg}_{(0)\,\mu\nu\sigma}(p,q,r) 
\Gamma^{\bar{c}c}_{(1)}(r) 
\label{3ptid}
\end{eqnarray}
where we recall a subscript $0$ on a Green's function means the tree 
contribution only. We note that we have set 
\begin{equation}
\Gamma_{\mu\nu}^{g}(p) ~=~ P_{\mu\nu}(p) ~+~ \Gamma^{gg}_{(1)}(p) P_{\mu\nu}(p)
\end{equation}
for the inverse propagator to one loop with $\Gamma^{\bar{c}c}_{(1)}$ 
indicating the one loop correction of the ghost leg of the final graph of 
Figure \ref{fig3pt}. For reference we note 
\begin{eqnarray}
\Gamma^{gg}_{(1)}(p) &=& \left[ \left[  
\frac{5}{3} C_A
- \frac{4}{3} \Nf T_F
+ \frac{1}{2} \xi C_A
\right] \frac{1}{\epsilon}
+ \frac{31}{9} C_A
- \frac{20}{9} \Nf T_F
- \xi C_A
+ \frac{1}{4} \xi^2 C_A
\right. \nonumber \\
&& \left. ~
+ \left[ 
\frac{188}{27} C_A
- \frac{112}{27} \Nf T_F
- 2 \xi C_A
+ \frac{1}{2} \xi^2 C_A
\right] \epsilon ~+~ O(\epsilon^2)
\right] g^2 
\end{eqnarray}
and
\begin{equation}
\Gamma^{\bar{c}c}_{(1)}(p) ~=~ 
\left[ \frac{1}{2 \epsilon} C_A 
+ \frac{1}{4 \epsilon} \xi C_A 
+ C_A
+ 2 C_A \epsilon
+ O(\epsilon^2)
\right] g^2 
\end{equation}
as well as 
\begin{equation}
\Gamma^{ggg}_{(0)\,\mu\nu\rho}(p,q,r) ~=~ 
i \left[ 
\eta_{\mu\nu} q_\sigma 
- \eta_{\mu\nu} p_\sigma
+ 2 \eta_{\mu\sigma} p_\nu
+ \eta_{\mu\sigma} q_\nu
- \eta_{\nu\sigma} p_\mu 
- 2 \eta_{\nu\sigma} q_\mu
\right] g 
\end{equation}
for the tree term of the $3$-point gluon vertex. We have included diagrams with
quarks in all our computations and their contributions are associated with the 
number of flavours $\Nf$ and Dynkin index $T_F$. However for the $3$-point 
identity (\ref{3ptid}) they primarily play a passive role in the verification 
by calculation.

{\begin{table}[ht]
\begin{center}
\begin{tabular}{|c||c|c|c|c|c|}
\hline
\rule{0pt}{12pt}
Entity & ${\cal P}^{g\bar{c}c}_{(1)}$ & ${\cal P}^{g\bar{c}c}_{(2)}$ & 
${\cal P}^{g\bar{c}c}_{(3)}$ & ${\cal P}^{g\bar{c}c}_{(4)}$ & 
${\cal P}^{g\bar{c}c}_{(5)}$ \\
\hline
\hline
\rule{0pt}{12pt}
$\frac{1}{\epsilon} \frac{}{}$ & 
$0$ & $0$ & $0$ & $0$ & $0$ \\
\rule{0pt}{12pt}
$\mathbb{Q}$ & 
$- \frac{3}{8} C_A$ & $- \frac{1}{6} C_A$ & 
$- \frac{1}{12} C_A$ & $- \frac{1}{12} C_A$ & $- \frac{5}{12} C_A$ \\
\rule{0pt}{12pt}
$\pi^2$ & 
$\frac{1}{54} C_A$ & $\frac{1}{6} C_A$ & $\frac{1}{16} C_A$ &
$\frac{2}{27} C_A$ & $\frac{1}{18} C_A$ \\
\rule{0pt}{12pt}
$\psi^\prime(\third)$ & 
$- \frac{1}{36} C_A$ & $- \frac{1}{4} C_A$ & $- \frac{1}{8} C_A$ & 
$- \frac{1}{12} C_A$ & $- \frac{1}{36} C_A$ \\
\hline
\hline
\rule{0pt}{12pt}
$\frac{1}{\epsilon} \frac{}{}$ & 
$0$ & $0$ & $0$ & $0$ & $0$ \\
\rule{0pt}{12pt}
$\mathbb{Q}$ &
$\frac{3}{8} C_A$ & $\frac{5}{12} C_A$ & 
$\frac{1}{12} C_A$ & $\frac{1}{12} C_A$ & $\frac{1}{6} C_A$ \\
\rule{0pt}{12pt}
$\pi^2$ & 
$- \frac{1}{54} C_A$ & $- \frac{1}{18} C_A$ & $- \frac{1}{12} C_A$ &
$- \frac{2}{27} C_A$ & $- \frac{1}{6} C_A$ \\
\rule{0pt}{12pt}
$\psi^\prime(\third)$ &
$\frac{1}{36} C_A$ & $\frac{1}{12} C_A$ & $\frac{1}{8} C_A$ & 
$\frac{1}{9} C_A$ & $\frac{1}{4} C_A$ \\
\hline
\hline
\rule{0pt}{12pt}
$\frac{1}{\epsilon} \frac{}{}$ & 
$\frac{4}{3} T_F \Nf - \frac{13}{6} C_A$ & 
$\frac{4}{3} T_F \Nf - \frac{13}{6} C_A$ & $0$ & $0$ & $0$ \\
\rule{0pt}{12pt}
$\mathbb{Q}$ & 
$\frac{20}{9} T_F \Nf - \frac{97}{36} C_A$ & 
$\frac{20}{9} T_F \Nf - \frac{97}{36} C_A$ & $0$ & $0$ & $0$ \\
\rule{0pt}{12pt}
$\pi^2$ & 
$0$ & $0$ & $0$ & $0$ & $0$ \\
\rule{0pt}{12pt}
$\psi^\prime(\third)$ & 
$0$ & $0$ & $0$ & $0$ & $0$ \\
\hline
\hline
\rule{0pt}{12pt}
$\frac{1}{\epsilon} \frac{}{}$ & 
$- \frac{4}{3} T_F \Nf + \frac{13}{6} C_A$ & $0$ & $0$ & $0$ & 
$- \frac{4}{3} T_F \Nf + \frac{13}{6} C_A$ \\
\rule{0pt}{12pt}
$\mathbb{Q}$ & 
$- \frac{20}{9} T_F \Nf + \frac{97}{36} C_A$ & $0$ & $0$ & $0$ & 
$- \frac{20}{9} T_F \Nf + \frac{97}{36} C_A$  \\
\rule{0pt}{12pt}
$\pi^2$ & 
$0$ & $0$ & $0$ & $0$ & $0$ \\
\rule{0pt}{12pt}
$\psi^\prime(\third)$ & 
$0$ & $0$ & $0$ & $0$ & $0$ \\
\hline
\hline
\rule{0pt}{12pt}
$\frac{1}{\epsilon} \frac{}{}$ & 
$0$ & $\frac{3}{4} C_A$ & $0$ & $0$ & $- \frac{3}{4} C_A$ \\
\rule{0pt}{12pt}
$\mathbb{Q}$ & 
$0$ & $C_A$ & $0$ & $0$ & $- C_A$ \\
\rule{0pt}{12pt}
$\pi^2$ & 
$0$ & $0$ & $0$ & $0$ & $0$ \\
\rule{0pt}{12pt}
$\psi^\prime(\third)$ &
$0$ & $0$ & $0$ & $0$ & $0$ \\
\hline
\end{tabular}
\end{center}
\begin{center}
{Table $1$. Coefficients of each of the tensors and different structures for
each of the five terms on the right hand side of (\ref{3ptid}) in the Landau 
gauge.}
\end{center}
\end{table}}

To illustrate how each of the various terms of (\ref{3ptid}) conspire together
to satisfy the identity we have provided the values for each of the terms on 
the right hand side of (\ref{3ptid}) in Table $1$. We do this for the
completely symmetric point for simplicity here purely due to the cumbersome
expressions for the fully off-shell case. In Table $1$ there are five separate 
sections which correspond to the respective terms of (\ref{3ptid}) in order. 
Within each graph we have divided the contributions by the respective
structures which can appear in the expression. These correspond respectively
to the residue of the simple pole in $\epsilon$, the rational finite part and 
the coefficients of $\pi^2$ and $\psi^\prime(\third)$. As each graph was
decomposed into the Lorentz basis there are five columns corresponding to the
basis of the factored ghost-gluon vertex. Therefore Table $1$ compactly
summarizes all contributions to the right hand side of (\ref{3ptid}). The
corresponding coefficients of the left hand side of the identity are given in
Table $2$ in the same notation. Therefore it is a straightforward exercise to
sum the respective coefficients for each structure and tensor in Table $1$ and
see that they completely tally with the corresponding entries in Table $2$. It
is worth noting that this represents a check of the Slavnov-Taylor identity
derived using the methods of \cite{15,16,17} for a non-exceptional momentum 
configuration. No external momenta have been nullified. We have repeated the
same check for the completely off-shell case for $x$ and $y$ not restricted to
unity and found the same total consistency of (\ref{3ptid}). This is 
non-trivial and to give an indication of the nature of the functions involved
in this case we note that the coefficient of ${\cal P}^{g\bar{c}c}_{(1)}$ for 
the first term on the right hand side of (\ref{3ptid}) is
\begin{eqnarray} 
&& \left[
\frac{1}{4} y \ln(x)
- \frac{3}{8} y
- \frac{1}{2} y \ln(y)
+ \left[
\frac{1}{2} y^2
- \frac{3}{8} y
- \frac{1}{4} x y
\right] \Phi_1(x,y) 
\right. \nonumber \\
&& \left.
+ \left[
\left[
\frac{1}{8} y
- \frac{1}{4} y^2
+ \frac{1}{8} y^3
+ \frac{3}{8} x y
- \frac{1}{8} x y^2
\right] \ln(x)
+ \left[
\frac{1}{4} y
- \frac{1}{8} y^2
- \frac{1}{8} y^3
- \frac{1}{4} x y
+ \frac{1}{8} x y^2
\right] \ln(y)
\right. \right.
\nonumber \\
&& \left. \left. + \left[
\frac{1}{4} y
- \frac{1}{2} y^2
+ \frac{1}{4} y^3
- \frac{1}{4} x y
\right] \Phi_1(x,y) 
\right] \frac{1}{\Delta_G}
\right] C_A 
\label{ccg11}
\end{eqnarray} 
which is considerably more involved than the three coefficients in the first
column of Table $1$ corresponding to this graph. At the symmetric point 
(\ref{ccg11}) reduces to the corresponding entry where the function 
$\Phi_1(x,y)$ contains the polylogarithm function $\mbox{Li}_n(z)$ and is 
defined by, \cite{32},
\begin{equation}
\Phi_1(x,y) ~=~ \frac{1}{\lambda} \left[ 2 \mbox{Li}_2(-\rho x)
+ 2 \mbox{Li}_2(-\rho y)
+ \ln \left( \frac{y}{x} \right)
\ln \left( \frac{(1+\rho y)}{(1+\rho x)} \right)
+ \ln(\rho x) \ln(\rho y) + \frac{\pi^2}{3} \right]
\end{equation}
where
\begin{equation}
\rho(x,y) ~=~ \frac{2}{[1-x-y+\lambda(x,y)]} ~~~,~~~
\lambda(x,y) ~=~ \sqrt{\Delta_G} ~.
\end{equation}
The more involved $x$ and $y$ dependence would make the extension of Table $1$
to the off-shell case large but the data file contains the details of the full 
off-shell case for arbitrary gauge. However we confirm that the sum of the 
graphs on the right side of (\ref{3ptid}) fully agree with the expression for  
$\frac{r^\sigma}{r^2} \Gamma^{ggg}_{\mu\nu\sigma}(p,q,r)$ for non-unit $x$ and
$y$. We note that in addition to $\Phi_1(x,y)$ the $O(\epsilon)$ correction to
one loop master triangle graph is required for several Green's function
contributing to the $4$-point identity. Therefore as this is the appropriate
place to note this we record that the $O(\epsilon)$ term of the triangle master
is  
\begin{eqnarray}
\Psi_1(x,y) &=& -~ \frac{1}{\lambda} \left[ 
4 \mbox{Li}_3 \left( - \frac{\rho x(1+\rho y)}{(1-\rho^2 xy)} \right) 
+ 4 \mbox{Li}_3 \left( - \frac{\rho y(1+\rho x)}{(1-\rho^2 xy)} \right) 
- 4 \mbox{Li}_3 \left( - \frac{xy \rho^2}{(1-\rho^2 xy)} \right) 
\right. \nonumber \\
&& \left. ~~~~~~~
+ 2 \mbox{Li}_3 \left( \frac{x \rho(1+\rho y)}{(1+\rho x)} \right) 
+ 2 \mbox{Li}_3 \left( \frac{y \rho(1+\rho x)}{(1+\rho y)} \right) 
- 2 \mbox{Li}_3 ( \rho^2 xy ) 
- 2 \zeta_3
\right. \nonumber \\
&& \left. ~~~~~~~
- 2 \ln (y) \mbox{Li}_2 \left( \frac{x \rho(1+\rho y)}{(1+\rho x)} \right) 
- 2 \ln (x) \mbox{Li}_2 \left( \frac{y \rho(1+\rho x)}{(1+\rho y)} \right) 
- \frac{2}{3} \ln^3 \left( 1-\rho^2 xy \right)
\right. \nonumber \\
&& \left. ~~~~~~~
+ \frac{2}{3} \ln^3 \left( 1+\rho x \right)
+ \frac{2}{3} \ln^3 \left( 1+\rho y \right)
+ 2 \ln(\rho) \ln^2 \left( 1-\rho^2 x y \right)
\right. \nonumber \\
&& \left. ~~~~~~~
- 2 \ln(1-\rho^2 xy ) \left[ \ln(\rho x) \ln(\rho y)
+ \ln \left( \frac{y}{x} \right)
\ln \left( \frac{(1+\rho y)}{(1+\rho x)} \right)
\right. \right. \nonumber \\
&& \left. \left. ~~~~~~~~~~~~~~~~~~~~~~~~~~~~~~
+ 2 \ln(1+\rho x) \ln(1+\rho y) + \frac{\pi^2}{3} \right] 
\right. \nonumber \\
&& \left. ~~~~~~~
+ \frac{1}{2} \ln \left( xy \rho^2 \right) \left[ \ln(\rho x) \ln(\rho y)
+ \ln \left( \frac{y}{x} \right)
\ln \left( \frac{(1+\rho y)}{(1+\rho x)} \right)
- \ln^2 \left( \frac{(1+ \rho x)}{(1+\rho y)} \right) 
\right. \right. \nonumber \\
&& \left. \left. ~~~~~~~~~~~~~~~~~~~~~~~~~~~
+ \frac{2\pi^2}{3} 
\right] \right] ~. 
\end{eqnarray}

{\begin{table}[ht]
\begin{center}
\begin{tabular}{|c||c|c|c|c|c|}
\hline
\rule{0pt}{12pt}
Entity & ${\cal P}^{g\bar{c}c}_{(1)}$ & ${\cal P}^{g\bar{c}c}_{(2)}$ & 
${\cal P}^{g\bar{c}c}_{(3)}$ & ${\cal P}^{g\bar{c}c}_{(4)}$ & 
${\cal P}^{g\bar{c}c}_{(5)}$ \\
\hline
\hline
\rule{0pt}{12pt}
$\frac{1}{\epsilon} \frac{}{}$ & 
$0$ & $\frac{4}{3} T_F \Nf - \frac{13}{6} C_A$ & $0$ & $0$ & 
$- \frac{4}{3} T_F \Nf + \frac{13}{6} C_A$ \\
\rule{0pt}{12pt}
$\mathbb{Q}$ & 
$0$ & $\frac{20}{9} T_F \Nf - \frac{13}{9} C_A$ & $0$ & $0$ & 
$- \frac{20}{9} T_F \Nf + \frac{13}{9} C_A$ \\
\rule{0pt}{12pt}
$\pi^2$ & 
$0$ & $\frac{1}{9} C_A$ & $0$ & $0$ & $- \frac{1}{9} C_A$ \\
\rule{0pt}{12pt}
$\psi^\prime(\third)$ & 
$0$ & $- \frac{1}{6} C_A$ & $0$ & $0$ & $\frac{1}{6} C_A$ \\
\hline
\end{tabular}
\end{center}
\begin{center}
{Table $2$. Coefficients of each of the tensors and different structures for
the left hand side of (\ref{3ptid}) in the Landau gauge.}
\end{center}
\end{table}}

\sect{Quartic gluon vertex.}

We now turn to the examination of the $4$-point identity which was considered
in \cite{3,5,11} and that derived using algebraic and diagrammatic methods 
which is illustrated in Figure \ref{fig4pt}. It relates the purely gluonic 
$4$-point vertex function to three ghost-gluon boxes as well as the $4$-point 
functions built from reduced $3$-point ghost-gluon functions. We have labelled 
the respective orientations with $C$ and $D$ as they differ from the previous 
ones and illustrated their definitions in Figure \ref{figaccprex} for clarity. 
The graphical definitions of the $4$-point terms are provided in Figure 
\ref{figaaaapr} which includes the three orientations of the reduced $4$-point 
ghost-gluon functions where one external ghost leg corresponds to the reduced 
$3$-point ghost-gluon vertex. For reference we note that the quartic gluon 
vertex is defined by 
\begin{equation}
\left\langle A^a_\mu(p) A^b_\nu(q) A^c_\sigma(r) A^d_\rho(s) \right\rangle ~=~ 
\Gamma^{gggg\,abcd}_{\mu \nu \sigma \rho}(p,q,r,s) ~.
\end{equation}
Unlike its $3$-point counterpart we cannot factor off a common group theory
structure. This is because even at one loop there are a large number of 
different combinations of the structure constants and unit matrix in colour
space to produce symmetric rank $4$ colour tensors. This was also apparent in 
the early work of \cite{11} and had to be taken into account in \cite{21} as 
well. In Appendix B we have summarized our algorithm for dealing with aspects 
of the group theory issues which is based on \cite{21}. Though we note that 
throughout this section alone we restrict ourselves to the $SU(\Nc)$ group 
rather than the general Lie group considered previously. The $4$-point identity
of Figure \ref{figaaaapr} corresponds to 
\begin{eqnarray}
\Gamma^{gggg\,abcd}_{\mu\nu\sigma\rho}(p,q,r,s) s^\rho &=&
\Gamma^{c\bar{c}gg\,abcd}_{A\,\nu\sigma\lambda}(p,q,r,s) 
P^\lambda_{~\mu}(p) ~+~
\Gamma^{c\bar{c}gg\,abcd}_{B\,\mu\sigma\lambda}(p,q,r,s) P^\lambda_{~\nu}(q)
\nonumber \\
&& +~ \Gamma^{c\bar{c}gg\,abcd}_{C\,\mu\nu\lambda}(p,q,r,s) 
P^\lambda_{~\sigma}(r) 
\nonumber \\
&& +~
\Gamma^{g\bar{c}c\,dbe}_{D\,\nu\lambda}(s,q,-s-q) 
\Gamma^{ggg\,ace}_{\mu\sigma\lambda}(p,r,-p-r) ((p+r)^2)^{-\epsilon} 
\nonumber \\
&& +~
\Gamma^{g\bar{c}c\,cde}_{C\,\sigma\lambda}(r,s,-r-s) 
\Gamma^{ggg\,abe}_{\mu\nu\lambda}(p,q,-p-q) ((p+q)^2)^{-\epsilon} 
\nonumber \\
&& +~
\Gamma^{g\bar{c}c\,dae}_{D\,\mu\lambda}(s,p,-s-p) 
\Gamma^{ggg\,bce}_{\nu\sigma\lambda}(q,r,-q-r) ((q+r)^2)^{-\epsilon} 
\nonumber \\
&& +~ \Gamma^{gggg\,abcd}_{(0)\,\mu\nu\sigma\rho}(p,q,r,s) s^\rho \, 
\Gamma^{\bar{c}c}_{(1)}(s) 
\label{4ptid}
\end{eqnarray}
or to (\ref{hoch4pt}) with the Lorentz and colour indices included for the 
practical task of its evaluation. Due to the presence of two $3$-point vertices
appearing in graphs with a bridge we have temporarily reintroduced the colour 
indices in those vertex functions to assist with the placement of the structure
constants in an evaluation.

\vspace{0.3cm}
{\begin{figure}[ht]
\begin{center}
\includegraphics[width=11.0cm,height=5.5cm]{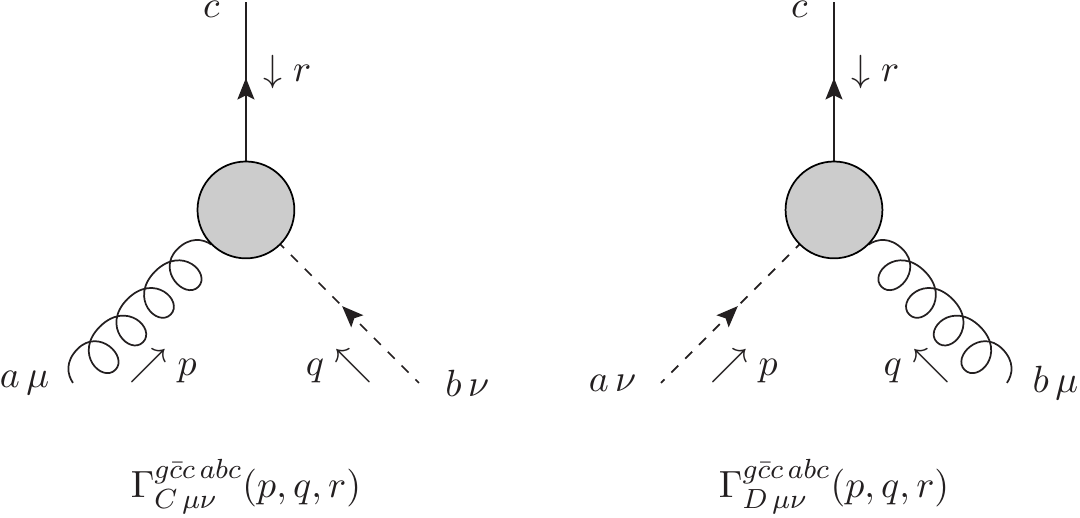}
\end{center}
\caption{Extra ghost-gluon $3$-point function configurations for $4$-point 
identity.}
\label{figaccprex}
\end{figure}}

To study (\ref{4ptid}) we will restrict to the fully symmetric point setup of
\cite{11} but using the notation of \cite{21}. For a $4$-point function the
symmetric point is defined by the following relations between the external
momenta 
\begin{equation}
p^2 ~=~ q^2 ~=~ r^2 ~=~ -~ \mu^2 ~~,~~
pq ~=~ pr ~=~ qr ~=~ \frac{1}{3} \mu^2
\label{symm4pt}
\end{equation}
where we take $s$ as the dependent momentum since
\begin{equation}
p ~+~ q ~+~ r ~+~ s ~=~ 0 ~.
\end{equation}
The Mandelstam variables are explicitly defined by 
\begin{equation}
\bar{s} ~=~ \frac{1}{2} ( p + q )^2 ~~,~~ 
\bar{t} ~=~ \frac{1}{2} ( q + r )^2 ~~,~~
\bar{u} ~=~ \frac{1}{2} ( p + r )^2
\label{mandvar} 
\end{equation}
then take the values
\begin{equation}
\bar{s} ~=~ \bar{t} ~=~ \bar{u} ~=~ -~ \frac{4}{3} \mu^2 ~. 
\label{mandval} 
\end{equation}
Here $\mu$ is the overall mass scale to which all the external momenta relate
to. In the previous section we considered the completely off-shell $3$-point
identity and it is apparent in Figure \ref{figaaaapr} that these functions play
a role. However at the $4$-point symmetric point it is important to realise
that in bolting two $3$-point functions together each of these functions are 
{\em not} at the $3$-point symmetric point. With (\ref{symm4pt}) the bridging 
momentum corresponds to one of the Mandelstam variables (\ref{mandvar}). 

Aside from the various different colour channels which are present in the
$4$-point functions there is a larger number of Lorentz tensors for each of
the Green's functions of Figure \ref{figaaaapr}. For instance 
$\Gamma^{gggg\,abcd}_{\mu\nu\sigma\rho}(p,q,r,s)$ has $138$ such tensors,
\cite{21}, which reduces to $36$ when contracted with $s^\rho$. These match
the same $36$ possibilities for the decomposition of 
$\Gamma^{c\bar{c}gg\,abcd}_{\nu\sigma\lambda}(p,q,r,s)$ into its tensor basis. 
As this Green's function was not computed in \cite{22} we need to evaluate it 
here for the three different routings. To do so we follow the same procedure as 
(\ref{accpradecomp}). The tensor basis for each case is structurally the same 
differing only in the rotation of the Lorentz indices with respect to the 
various external legs. In Appendix A we have given the tensor basis as well as 
the projection matrix for the $4$-point ghost-gluon function at the symmetric 
point. For a more general off-shell setup the projection matrix method is not 
appropriate to use given the large dependence on the kinematic variables which 
would be present. Aside from the Mandelstam variables the other variables in 
that instance would be the ratios of the three dependent external momenta, akin
to $x$ and $y$ of (\ref{3ptvar}), as well as one overall mass scale which would
be $\mu^2$. To be clear we decompose the ghost-gluon $4$-point function of the 
identity via
\begin{equation} 
\left\langle A^a_\mu(p) A^b_\nu(q) c^c(r) \bar{c}_\sigma^d(s) 
\right\rangle ~=~ \Gamma^{c\bar{c}gg\,abcd}_{\mu\nu\sigma}(p,q,r,s) ~=~
\sum_{k=1}^{36} {\cal P}^{c\bar{c}gg}_{(k) \mu\nu\sigma}(p,q,r) \,
\Sigma^{c\bar{c}gg\,abcd}_{(k)}(p,q,r)
\label{aaccprdecomp}
\end{equation} 
using the same basic notation as earlier. The full Lorentz tensor basis is 
given in Appendix A together with the elements of the projection matrix 
${\cal M}^{c\bar{c}gg}$ at the symmetric point (\ref{symm4pt}). For each of the
three orientations there are $7$ Feynman graphs at one loop none of which 
involve quarks. Again the method we employ to evaluate the scalar amplitudes of
$\Gamma^{c\bar{c}gg\,abcd}_{\mu\nu\sigma}(p,q,r,s)$ is the same as that for the
$3$-point case in that we use the Laporta algorithm after the scalar amplitudes
have been isolated by the projection method. One of the main differences in 
computing $\Gamma^{gggg\,abcd}_{\mu\nu\sigma\rho}(p,q,r,s)$ and the various 
orientations of $\Gamma^{c\bar{c}gg\,abcd}_{\mu\nu\sigma}(p,q,r,s)$ is that 
there is more than one possible rank $4$ colour group tensor which can appear 
at one loop. For a $3$-point function involving only ghosts and gluons there is
only one possible rank $3$ colour tensor. For the $4$-point functions we have 
to be careful in our colour tensor basis choice and have retained that which 
was used in \cite{21,22}. The technical details of this have been relegated to 
Appendix B but we note that we do not use a method of projection similar to 
that for the Lorentz structure. Instead we map the colour combinations into our
choice of basis tensors. For completeness this is
\begin{equation}
\left\{ \delta^{ab} \delta^{cd}, \delta^{ac} \delta^{bd}, 
\delta^{ad} \delta^{bc}, f^{abe} f^{cde}, f^{ace} f^{bde}, d_A^{abcd}, 
d_F^{abcd} \right\}
\end{equation}
where the fully symmetric rank $4$ tensors $d_A^{abcd}$ and $d_F^{abcd}$ are
defined in (\ref{defrank4}) and their properties discussed at length in 
\cite{36}. We have used the Jacobi identity to ensure there are only two 
independent rank $4$ combinations of the product of two structure constants. 
For illustration we have provided the expressions for 
$\Gamma^{gggg\,abcd}_{\mu\nu\sigma\rho}(p,q,r,s)$ and 
$\Gamma^{c\bar{c}gg\,abcd}_{A\,\mu\nu\sigma}(p,q,r,s)$ in the Landau gauge at 
the fully symmetric point in Appendix C. Full explicit expressions for each of 
the graphs in Figure \ref{fig4pt} in an arbitrary linear covariant gauge are 
given in the attached data file. 

\vspace{0.3cm}
{\begin{figure}[ht]
\begin{center}
\includegraphics[width=10.0cm,height=10.0cm]{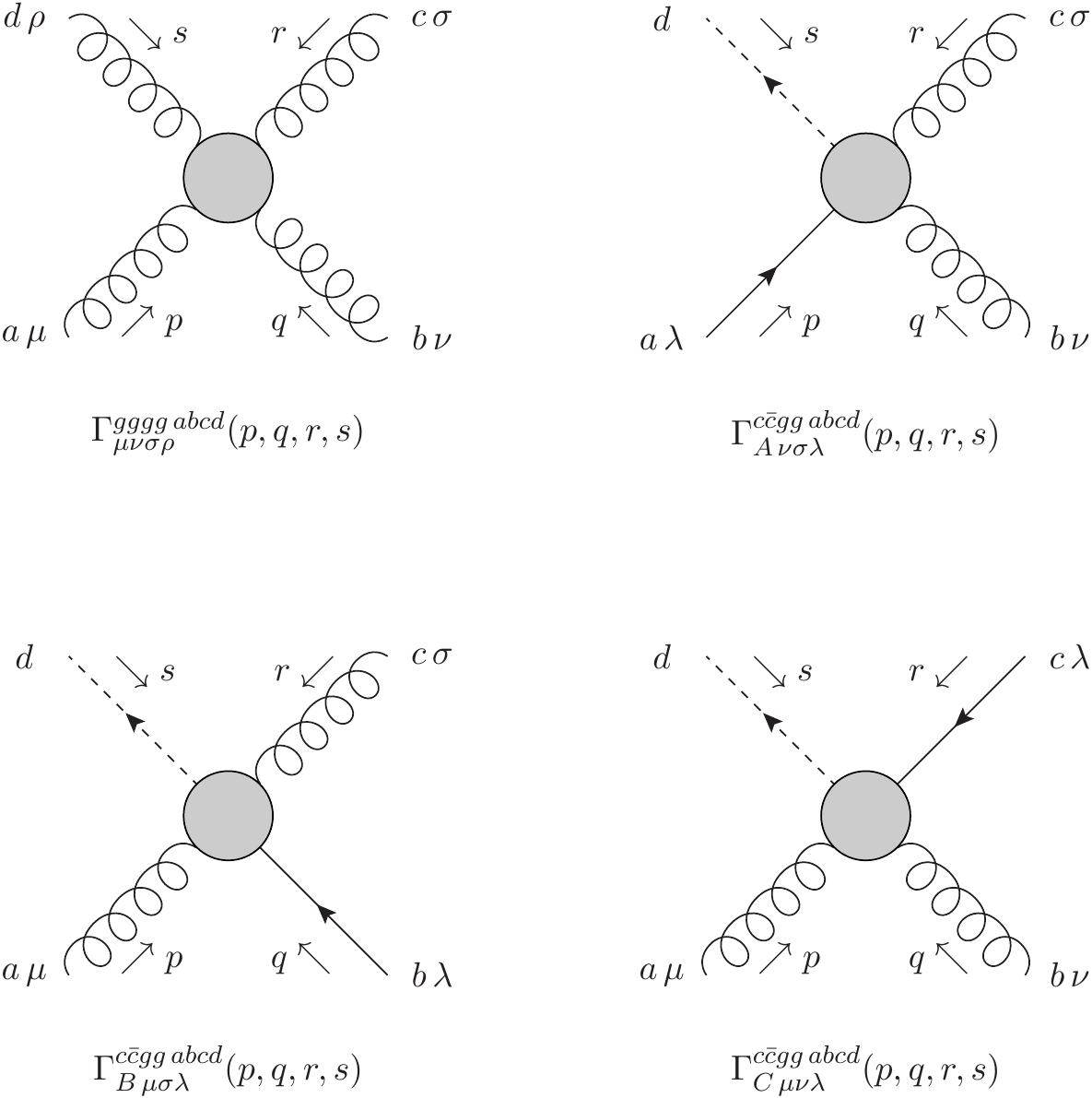}
\end{center}
\caption{Basic one particle irreducible Green's functions for $4$-point 
function.}
\label{figaaaapr}
\end{figure}}

\vspace{0.3cm}
{\begin{figure}[ht]
\begin{center}
\includegraphics[width=9.0cm,height=7.0cm]{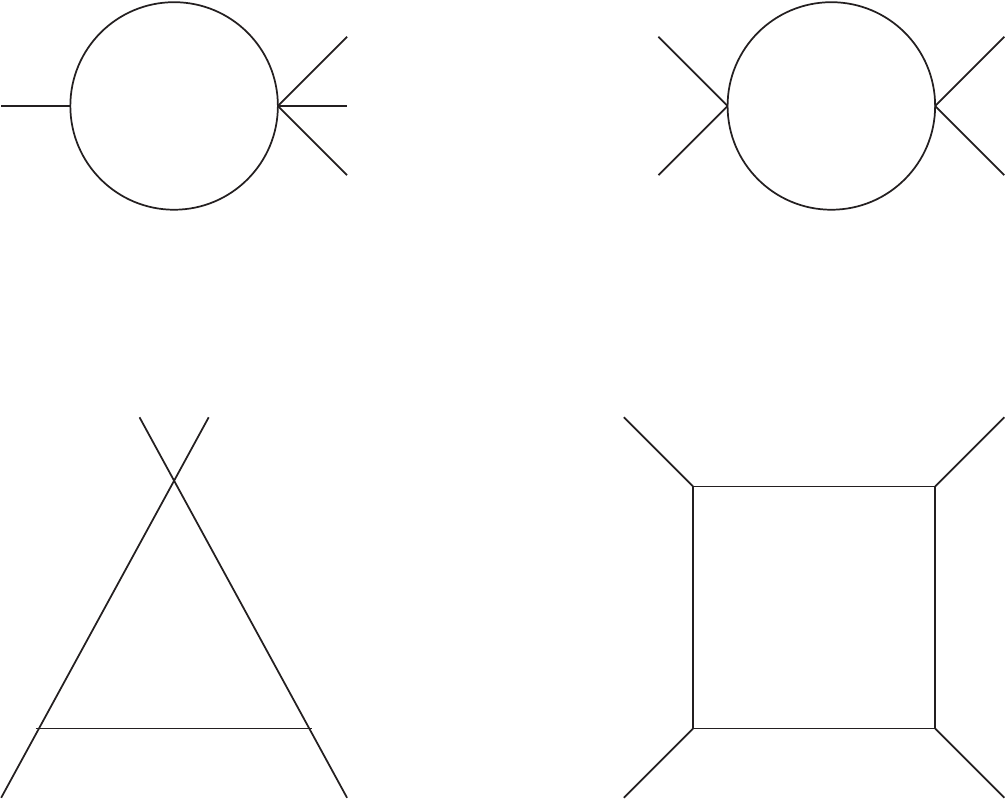}
\end{center}
\caption{Master integrals for reduction of $4$-point function.}
\label{figmast4}
\end{figure}}

Having discussed the technical issues around computing the individual graphs
contributing to (\ref{4ptid}) we now turn to demonstrating that the identity is
satisfied at the symmetric point (\ref{symm4pt}). Compared to the $3$-point
case this is much more involved because of the structure of the Green's
function. First there are three colour and $36$ Lorentz tensors in the basis
for each graph. On top of this there are four different numerological 
structures in each tensor coefficient aside from the pole term in $\epsilon$.
These structures are the pure rational piece, $\ln(\fourthirds)$ and two
specific functions involving dilogarithms. These are
$\Phi_1(\threequarters,\threequarters)$ and 
$\Phi_1(\ninesixteenths,\ninesixteenths)$. This gives a large number of overall
terms which have to sum to match both sides of (\ref{4ptid}) exactly. If we 
dissect the origin of $\Phi_1(\threequarters,\threequarters)$ and 
$\Phi_1(\ninesixteenths,\ninesixteenths)$ more carefully we can understand how
part of this cancellation proceeds. The arguments of both functions reflect
the kinematics of the underlying master integral deriving from the application
of the Laporta algorithm. For instance the completely off-shell box fully
symmetric point master integral of Figure \ref{figmast4} is given by, 
\cite{37,38},
\begin{equation}
\Phi_1 \left( \frac{p^2 r^2}{(p+q)^2 (q+r)^2}, \frac{q^2 s^2}{(p+q)^2 (q+r)^2} 
\right)
\end{equation}
which depend on the Mandelstam variables. Using their values at the symmetric
point, (\ref{mandval}), determines the origin of 
$\Phi_1(\ninesixteenths,\ninesixteenths)$. Therefore in (\ref{4ptid}) the only
terms where such a function can arise is in 1PI $4$-point functions which is 
the first three terms of the right hand side as well as the left side. The 
other structure $\Phi_1(\threequarters,\threequarters)$ occurs in all bar the 
last term on the right hand side of (\ref{4ptid}). However it has different 
origins in the three $4$-point functions and the three graphs of Figure 
\ref{fig4pt} where there is a bridging propagator between two $3$-point
functions. For instance the box graph of Figure \ref{figmast4} is also one
of the three box topologies in the integral family which is the starting point 
of the Laporta algorithm. Within the integration by parts routine various 
masters can be deduced by taking one of the integral family topologies and 
removing lines or propagators in such a way as not to produce a zero sector 
integral. Doing this for the box graph of Figure \ref{figmast4} produces the 
three remaining graphs of that Figure. Given this 
$\Phi_1(\threequarters,\threequarters)$ corresponds to the triangle graph.
We have deliberately retained the two external lines of the top apex to
indicate the origin of this master. However both lines reflect that the 
momentum flowing out of that point is the sum of the two incoming momenta.
Therefore from (\ref{symm4pt}) one does not have a triangle master at the fully
symmetric point of the $3$-point functions of the previous section. Instead the
variable for the top apex momentum is one of the three Mandelstam variables of 
(\ref{mandval}). For the connected graphs of Figure \ref{fig4pt} each $3$-point 
vertex function is likewise evaluated with the connecting vertex momentum 
taking the same Mandelstam variable to produce terms involving
$\Phi_1(\threequarters,\threequarters)$. Finally the remaining two bubble
topologies of Figure \ref{figmast4} give different values for similar reasons. 
This is because upon reduction from the box the momentum flow across the left 
bubble graph is $p$ for instance but $(p+q)$ say for the second. The latter 
involves a Mandelstam variable and also is one source of the $\ln(\fourthirds)$ 
structure in the final expressions.

In order to give a flavour of how these different structures are assembled to
ensure the Slavnov-Taylor identity is satisfied at the fully symmetric point we
have repeated the approach used for the $3$-point case. In other words we have 
provided a breakdown of the terms contributing to graphs in (\ref{4ptid}) in 
Tables $3$ to $6$. However given the large number of different tensor
structures, both colour and Lorentz, we focus on a representative selection of
each. In each of these tables the coefficients of the various structures are
given for each graph. We note not every graph has simple poles in $\epsilon$
and only record data for them where they occur. In each of the Tables $3$ to 
$6$ we focus on $5$ of the $36$ different Lorentz tensors noting for instance
in Table $3$ only $5$ graphs on the right side of (\ref{4ptid}) have
contributions for these tensors. For a different choice of Lorentz tensors
other graphs would contribute. To differentiate between the colour structures 
Tables $3$ and $4$ correspond to the colour tensor $f_4^{abcd}$ while Tables 
$5$ and $6$ relate to the symmetric tensor $d_A^{abcd}$ where both these 
tensors are defined in (\ref{ten4def}). Similar to before the respective terms 
in Table $3$ sum to the values give in Table $4$ which are the calculated 
values of the left side of (\ref{4ptid}). For the case of $d_A^{abcd}$ the 
respective Tables are $5$ and $6$ and we note that for the former table only 
the first three terms on the right hand side of (\ref{4ptid}) can have terms 
involving this colour tensor. As before we confirm that adding all the 
contributions from the right hand side of (\ref{4ptid}) gives precisely the 
same expression for fully agree with the expression for 
$s^\rho \Gamma^{gggg\,abcd}_{\mu\nu\sigma\rho}(p,q,r,s)$ for the configuration 
(\ref{symm4pt}) in all Lorentz and colour channels.

{\begin{table}[ht]
\begin{center}
\begin{tabular}{|c||c|c|c|c|c|}
\hline
\rule{0pt}{12pt}
Entity & ${\cal P}^{c\bar{c}gg}_{(1)}$ & ${\cal P}^{c\bar{c}gg}_{(10)}$ & 
${\cal P}^{c\bar{c}gg}_{(11)}$ &
${\cal P}^{c\bar{c}gg}_{(16)}$ & ${\cal P}^{c\bar{c}gg}_{(31)}$ \\
\hline
\hline
\rule{0pt}{12pt}
$\mathbb{Q}$ & 
$- \frac{63}{1280} \Nc$ & $- \frac{1609}{30720} \Nc$ & 
$- \frac{4151}{15360} \Nc$ & $\frac{37}{360} \Nc$ &
$- \frac{453}{1280} \Nc$ \\
\rule{0pt}{12pt}
$\ln(\fourthirds)$ &
$\frac{2821}{25600} \Nc$ & $- \frac{183663}{204800} \Nc$ & 
$- \frac{931}{1600} \Nc$ &
$\frac{113231}{307200} \Nc$ & $- \frac{36063}{102400} \Nc$ \\
\rule{0pt}{12pt}
$\tilde{\Phi}_1(\ninesixteenths)$ &
$\frac{33}{25600} \Nc$ & $\frac{18211}{204800} \Nc$ & 
$\frac{35329}{819200} \Nc$ & $- \frac{16307}{819200} \Nc$ & 
$- \frac{17427}{819200} \Nc$ \\
\rule{0pt}{12pt}
$\tilde{\Phi}_1(\threequarters)$ &
$\frac{101}{4096} \Nc$ & $- \frac{3983}{32768} \Nc$ & 
$- \frac{129}{4096} \Nc$ & $- \frac{111}{16384} \Nc$ & 
$\frac{2925}{16384} \Nc$ \\
\hline
\hline
\rule{0pt}{12pt}
$\mathbb{Q}$ & 
$\frac{23}{160} \Nc$ & $0$ & $- \frac{867}{5120} \Nc$ & 
$- \frac{257}{5760} \Nc$ & $- \frac{221}{5120} \Nc$ \\
\rule{0pt}{12pt}
$\ln(\fourthirds)$ &
$\frac{13}{25} \Nc$ & $0$ & $\frac{4791}{51200} \Nc$ & 
$- \frac{17929}{76800} \Nc$ &
$\frac{37989}{51200} \Nc$ \\
\rule{0pt}{12pt}
$\tilde{\Phi}_1(\ninesixteenths)$ &
$- \frac{1649}{25600} \Nc$ & $0$ & $\frac{8253}{819200} \Nc$ & 
$- \frac{9137}{204800} \Nc$ & $\frac{103707}{819200} \Nc$ \\
\rule{0pt}{12pt}
$\tilde{\Phi}_1(\threequarters)$ &
$\frac{1}{128} \Nc$ & $0$ & $\frac{477}{8192} \Nc$ & 
$- \frac{63}{4096} \Nc$ & $- \frac{1521}{8192} \Nc$ \\
\hline
\hline
\rule{0pt}{12pt}
$\mathbb{Q}$ & 
$- \frac{263}{3840} \Nc$ & $\frac{1609}{30720} \Nc$ & 
$\frac{9331}{30720} \Nc$ & $- \frac{59}{9216} \Nc$ &
$- \frac{1009}{5120} \Nc$ \\
\rule{0pt}{12pt}
$\ln(\fourthirds)$ &
$\frac{15661}{76800} \Nc$ & $\frac{183663}{204800} \Nc$ & 
$\frac{154731}{204800} \Nc$ &
$\frac{6341}{30720} \Nc$ & $- \frac{1101}{102400} \Nc$ \\
\rule{0pt}{12pt}
$\tilde{\Phi}_1(\ninesixteenths)$ &
$\frac{3503}{76800} \Nc$ & $- \frac{18211}{204800} \Nc$ & 
$- \frac{69073}{819200} \Nc$ & $- \frac{17859}{163840} \Nc$ & 
$\frac{7331}{51200} \Nc$ \\
\rule{0pt}{12pt}
$\tilde{\Phi}_1(\threequarters)$ &
$- \frac{337}{4096} \Nc$ & $\frac{3983}{32768} \Nc$ & 
$\frac{563}{32768} \Nc$ & $\frac{945}{8192} \Nc$ & 
$\frac{1715}{16384} \Nc$ \\
\hline
\hline
\rule{0pt}{12pt}
$\frac{1}{\epsilon}$ & 
$\frac{17\Nc}{12} - \frac{2\Nf}{3}$ & $0$ & $0$ & $0$ & $0$ \\
\rule{0pt}{12pt}
$\mathbb{Q}$ & 
$- \frac{\Nc}{72} - \frac{37\Nf}{36}$ & 
$\frac{63\Nc}{64} + \frac{3\Nf}{32}$ & 
$\frac{73\Nc}{64} - \frac{11\Nf}{32}$ & $0$ &
$- \frac{\Nc}{4}$ \\
\rule{0pt}{12pt}
$\ln(\fourthirds)$ &
$\frac{95\Nc}{96} - \frac{7\Nf}{48}$ &
$\frac{191\Nc}{256} - \frac{5\Nf}{128}$ & 
$- \frac{847\Nc}{256} + \frac{93\Nf}{128}$ &
$0$ & $- \frac{25\Nc}{16}$ \\
\rule{0pt}{12pt}
$\tilde{\Phi}_1(\ninesixteenths)$ &
$0$ & $0$ & $0$ & $0$ & $0$ \\
\rule{0pt}{12pt}
$\tilde{\Phi}_1(\threequarters)$ &
$- \frac{43\Nc}{128} + \frac{27\Nf}{64}$ & 
$\frac{327\Nc}{1024} - \frac{189\Nf}{512}$ & 
$\frac{681\Nc}{1024} - \frac{171\Nf}{512}$ & $0$ & 
$- \frac{9\Nc}{64}$ \\
\hline
\hline
\rule{0pt}{12pt}
$\frac{1}{\epsilon}$ & 
$- \frac{17\Nc}{12} + \frac{2\Nf}{3}$ & $0$ & $0$ & $0$ & $0$ \\
\rule{0pt}{12pt}
$\mathbb{Q}$ & 
$- \frac{23\Nc}{72} + \frac{37\Nf}{36}$ & 
$- \frac{63\Nc}{64} - \frac{3\Nf}{32}$ & 
$- \frac{83\Nc}{64} + \frac{11\Nf}{16}$ & 
$- \frac{91\Nc}{192} - \frac{\Nf}{16}$ & 
$- \frac{15\Nc}{32} + \frac{9\Nf}{8}$ \\ 
\rule{0pt}{12pt}
$\ln(\fourthirds)$ &
$\frac{455\Nc}{96} - \frac{53\Nf}{48}$ &
$- \frac{191\Nc}{256} + \frac{5\Nf}{128}$ & 
$- \frac{203\Nc}{256} + \frac{19\Nf}{64}$ &
$\frac{847\Nc}{256} - \frac{41\Nf}{64}$ &
$- \frac{531\Nc}{128} + \frac{41\Nf}{32}$ \\
\rule{0pt}{12pt}
$\tilde{\Phi}_1(\ninesixteenths)$ &
$0$ & $0$ & $0$ & $0$ & $0$ \\
\rule{0pt}{12pt}
$\tilde{\Phi}_1(\threequarters)$ &
$- \frac{15\Nc}{128} + \frac{15\Nf}{64}$ & 
$- \frac{327\Nc}{1024} + \frac{189\Nf}{512}$ & 
$- \frac{243\Nc}{1024} + \frac{27\Nf}{256}$ & 
$- \frac{297\Nc}{1024} + \frac{63\Nf}{256}$ & 
$\frac{261\Nc}{512} - \frac{63\Nf}{128}$ \\
\hline
\end{tabular}
\end{center}
\begin{center}
{Table $3$. Coefficients of selected tensors and different structures for the
first three, fifth and sixth terms on the right hand side of (\ref{4ptid}) for
the $f_4^{abcd}$ sector in the Landau gauge.}
\end{center}
\end{table}}

{\begin{table}[ht]
\begin{center}
\begin{tabular}{|c||c|c|c|c|c|}
\hline
\rule{0pt}{12pt}
Entity & ${\cal P}^{c\bar{c}gg}_{(1)}$ & ${\cal P}^{c\bar{c}gg}_{(10)}$ & 
${\cal P}^{c\bar{c}gg}_{(11)}$ &
${\cal P}^{c\bar{c}gg}_{(16)}$ & ${\cal P}^{c\bar{c}gg}_{(31)}$ \\
\hline
\hline
\rule{0pt}{12pt}
$\mathbb{Q}$ & 
$- \frac{59\Nc}{72} - \frac{5\Nf}{4}$ & 
$0$ &
$- \frac{2991\Nc}{10240} + \frac{11\Nf}{32}$ & 
$- \frac{1297\Nc}{3072} - \frac{\Nf}{16}$ & 
$- \frac{3361\Nc}{2560} + \frac{9\Nf}{8}$ \\ 
\rule{0pt}{12pt}
$\ln(\fourthirds)$ &
$\frac{8401\Nc}{1280} - \frac{5\Nf}{4}$ &
$0$ &
$- \frac{785273\Nc}{204800} + \frac{131\Nf}{128}$ &
$\frac{14951\Nc}{4096} - \frac{41\Nf}{64}$ &
$- \frac{272993\Nc}{51200} + \frac{41\Nf}{32}$ \\
\rule{0pt}{12pt}
$\tilde{\Phi}_1(\ninesixteenths)$ &
$- \frac{269\Nc}{15360}$ & 
$0$ &
$- \frac{25491\Nc}{819200}$ & 
$- \frac{2843\Nc}{16384}$ & 
$\frac{25447\Nc}{102400}$ \\
\rule{0pt}{12pt}
$\tilde{\Phi}_1(\threequarters)$ &
$- \frac{515\Nc}{1024} + \frac{3\Nf}{16}$ & 
$0$ &
$\frac{15455\Nc}{32768} - \frac{117\Nf}{512}$ & 
$- \frac{3225\Nc}{16384} + \frac{63\Nf}{256}$ & 
$\frac{3823\Nc}{8192} - \frac{63\Nf}{128}$ \\
\hline
\end{tabular}
\end{center}
\begin{center}
{Table $4$. Coefficients of selected tensors and different structures for the
left hand side of (\ref{4ptid}) for the $f_4^{abcd}$ sector in the Landau 
gauge.}
\end{center}
\end{table}}

{\begin{table}[ht]
\begin{center}
\begin{tabular}{|c||c|c|c|c|c|}
\hline
\rule{0pt}{12pt}
Entity & ${\cal P}^{c\bar{c}gg}_{(1)}$ & ${\cal P}^{c\bar{c}gg}_{(10)}$ & 
${\cal P}^{c\bar{c}gg}_{(11)}$ &
${\cal P}^{c\bar{c}gg}_{(16)}$ & ${\cal P}^{c\bar{c}gg}_{(31)}$ \\
\hline
\hline
\rule{0pt}{12pt}
$\mathbb{Q}$ & 
$\frac{81}{640}$ & $\frac{2119}{2560}$ & 
$\frac{1}{256}$ & $\frac{709}{3840}$ &
$\frac{207}{640}$ \\
\rule{0pt}{12pt}
$\ln(\fourthirds)$ &
$\frac{6909}{6400}$ & $\frac{123771}{51200}$ & $\frac{93951}{51200}$ &
$- \frac{15791}{25600}$ & $- \frac{2007}{2048}$ \\
\rule{0pt}{12pt}
$\tilde{\Phi}_1(\ninesixteenths)$ &
$\frac{134337}{102400}$ & $\frac{6603}{3200}$ & 
$\frac{345189}{409600}$ & $- \frac{14397}{102400}$ & 
$- \frac{111879}{81920}$ \\
\rule{0pt}{12pt}
$\tilde{\Phi}_1(\threequarters)$ &
$- \frac{285}{1024}$ & $- \frac{23493}{8192}$ & 
$- \frac{8469}{8192}$ & $- \frac{135}{4096}$ & $\frac{10125}{8192}$ \\
\hline
\hline
\rule{0pt}{12pt}
$\mathbb{Q}$ & 
$- \frac{21}{80}$ & $0$ & 
$\frac{297}{640}$ & $\frac{709}{3840}$ &
$\frac{1209}{2560}$ \\
\rule{0pt}{12pt}
$\ln(\fourthirds)$ &
$\frac{1437}{1600}$ & $0$ & $\frac{64863}{51200}$ &
$- \frac{15791}{25600}$ & $\frac{11277}{5120}$ \\
\rule{0pt}{12pt}
$\tilde{\Phi}_1(\ninesixteenths)$ &
$- \frac{2967}{12800}$ & $0$ & 
$\frac{245187}{409600}$ & $- \frac{14397}{102400}$ & 
$\frac{336609}{81920}$ \\
\rule{0pt}{12pt}
$\tilde{\Phi}_1(\threequarters)$ &
$\frac{57}{256}$ & $0$ & 
$- \frac{5805}{8192}$ & $- \frac{135}{4096}$ & $- \frac{22761}{4096}$ \\
\hline
\hline
$\mathbb{Q}$ & 
\rule{0pt}{12pt}
$\frac{7}{640}$ & $- \frac{2119}{2560}$ & 
$- \frac{931}{2560}$ & $- \frac{103}{768}$ &
$- \frac{453}{640}$ \\
\rule{0pt}{12pt}
$\ln(\fourthirds)$ &
$- \frac{1057}{6400}$ & $- \frac{123771}{51200}$ & $- \frac{14727}{12800}$ &
$- \frac{1411}{1280}$ & $- \frac{24129}{10240}$ \\
\rule{0pt}{12pt}
$\tilde{\Phi}_1(\ninesixteenths)$ &
$- \frac{79301}{102400}$ & $- \frac{6603}{3200}$ & 
$- \frac{599997}{409600}$ & $\frac{49509}{40960}$ & 
$- \frac{44049}{16384}$ \\
\rule{0pt}{12pt}
$\tilde{\Phi}_1(\threequarters)$ &
$\frac{1161}{1024}$ & $\frac{23493}{8192}$ & 
$\frac{2211}{1024}$ & $- \frac{171}{128}$ & $\frac{24927}{8192}$ \\
\hline
\end{tabular}
\end{center}
\begin{center}
{Table $5$. Coefficients of selected tensors and different structures for the
first three terms on the right hand side of (\ref{4ptid}) for the $d_A^{abcd}$ 
sector in the Landau gauge.}
\end{center}
\end{table}}

{\begin{table}[ht]
\begin{center}
\begin{tabular}{|c||c|c|c|c|c|}
\hline
\rule{0pt}{12pt}
Entity & ${\cal P}^{c\bar{c}gg}_{(1)}$ & ${\cal P}^{c\bar{c}gg}_{(10)}$ & 
${\cal P}^{c\bar{c}gg}_{(11)}$ &
${\cal P}^{c\bar{c}gg}_{(16)}$ & ${\cal P}^{c\bar{c}gg}_{(31)}$ \\
\hline
\hline
\rule{0pt}{12pt}
$\mathbb{Q}$ & 
$- \frac{1}{8}$ & $0$ & $\frac{267}{2560}$ & 
$\frac{301}{1280}$ & $\frac{45}{512}$ \\
\rule{0pt}{12pt}
$\ln(\fourthirds)$ &
$\frac{29}{16}$ & $0$ & $\frac{49953}{25600}$ & $- \frac{29901}{12800}$ &
$- \frac{1161}{1024}$ \\
\rule{0pt}{12pt}
$\tilde{\Phi}_1(\ninesixteenths)$ &
$- \frac{7}{8}$ & $0$ & 
$- \frac{9621}{409600}$ & $\frac{189957}{204800}$ & 
$\frac{879}{16384}$ \\
\rule{0pt}{12pt}
$\tilde{\Phi}_1(\threequarters)$ &
$\frac{69}{64}$ & $0$ & $\frac{1707}{4096}$ &
$- \frac{2871}{2048}$ & $- \frac{5235}{4096}$ \\
\hline
\end{tabular}
\end{center}
\begin{center}
{Table $6$. Coefficients of selected tensors and different structures for the
left hand side of (\ref{4ptid}) for the $d_A^{abcd}$ sector in the Landau 
gauge.}
\end{center}
\end{table}}

\sect{Discussion.}

The main aim of our analysis was to first generate relations between various
$3$- and $4$-point 1PI Green's functions in QCD when the gauge fixing was the
canonical linear covariant one, and then to check that they were identically
satisfied by explicit one loop computations. To derive these identities, we 
start from the contraction $p_i \cdot \Gamma^n_c$~$=$~$0$ given by gauge 
invariance, which is the basic Slavnov-Taylor identity. Here, $\Gamma^n_c$ can 
be any connected Green's function with $n$ external gluons. The connected 
Green's function $\Gamma^n_c$ above has the structure of an amputated connected 
$n$-point vertex function where each of the external legs is then dressed by a 
propagator function.
Contraction of leg $i$ with its external momentum $p_i$ kills the transverse
degrees of freedom in leg $i$. Remaining is a longitudinal ghost self-energy 
which dresses leg $i$ in a connected Green's function as a through-going 
longitudinal degree of freedom exiting at any leg $j$ transversally. Hence the 
structure of the dressed Green's function is consistently maintained when 
replacing an external gluon leg by a longitudinal degree of freedom followed 
through the connected function in all possible ways. See, for example, Figure 
\ref{ScST}. At one loop for $\Gamma^{ggg}$ and $\Gamma^{gggg}$ this resulted in
an extra graph in each case and their presence could be deduced by the 
systematic use of Hopf-algebraic and diagrammatic formalisms of 
\cite{1,3,15,16,17,anatomy}. Indeed the approach in \cite{anatomy} was 
instrumental in gaining previous insights into the structure of gauge fixed QED
and QCD, \cite{18,19}. The extra graphs in both identities involved the ghost 
self-energy appended to an external leg as expected and its absence would have 
invalidated each relation calculationally. In the connected Green's function 
version of the Slavnov-Taylor identities such additional graphs are 
automatically incorporated. However in certain applications the 1PI version of 
the identities may be more applicable and our derivations and examples 
therefore crucially emphasize that one has to be careful in applying the 
correct relation. The fact that we check both our examples for non-exceptional 
momenta configurations, rather than nullifying an external leg as is carried 
out in some verifications, represents a robust check and circumvents any 
potential infrared issues. This is particularly the case for the $4$-point 
function given the large number of colour and Lorentz tensors that are present 
in each of the contributing graphs.

\vspace{0.3cm}
{\begin{figure}
\begin{center}
\includegraphics[width=14.5cm,height=5.5cm]{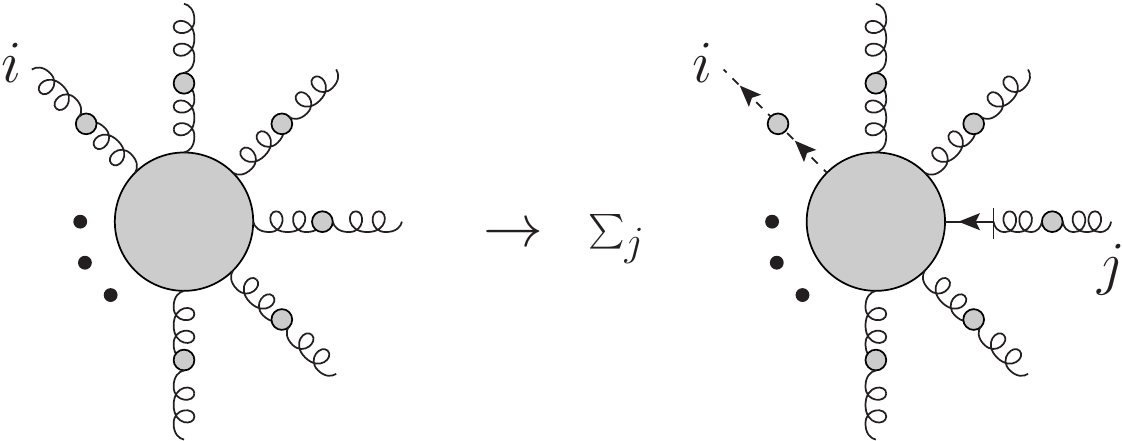}
\end{center}
\caption{Self-consistent Slavnov-Taylor identity. A connected $n$-gluon vertex 
function with $n$ dressed external gluon propagators when contracted at leg $i$
becomes a dressed vertex for a connected vertex function with $n-2$ external 
gluons and one through-going ghost line, providing one external ghost 
propagator and one transversal external leg and $n-2$ further external dressed 
gluon propagators.}
\label{ScST}
\end{figure}}

In light of these there are several natural avenues to pursue. One obvious one
is to first derive general relations for other gluon $n$-point functions as 
well as to extend our calculations to the two loop case. For the $3$-point 
identity this would be the first place to start since the full off-shell vertex
functions of QCD are known, \cite{26}. So the necessary computational 
tools are in place to determine the two loop corrections to 
$\Gamma^{g\bar{c}c}_{\mu\nu}(p,q,r)$ that would be needed. This would also 
provide the forum to study the effect and relation the identities we have
derived have on the renormalization of QCD in kinematical schemes such as the
momentum subtraction one of \cite{7,8}. Equally at a practical level the 
current use of the Dyson-Schwinger methods to probe the infrared behaviour of 
$2$-point and vertex functions in QCD, 
\cite{12,13,14,ver1,ver2,ver3,ver4,ver5,ver6,ver7,ver8,ver9,ver10,ver11,ver12,
ver13},
relies on relations between different $n$-point functions. Therefore 
constructing the identity for the $5$-point gluon function, for instance, could
provide a useful off-shell consistency check. Another direction that has been 
followed in recent years is to consider nonlinear gauges such as the 
Curci-Ferrari, \cite{39}, and maximal abelian gauges, \cite{40,41,42}. From a 
structural point of view the former gauge would be the starting place to 
understand the subtleties of the nonlinear aspect of such gauges. The former 
gauge differs from the linear covariant gauge only in having an asymmetric 
ghost-gluon vertex as well as a quartic ghost interaction. While the latter is 
a valid term in a renormalizable gauge theory, BRST symmetry excludes it in the
linear covariant case. Therefore given the insights we have found here we aim 
to study some of these potential new directions in future work. 

\vspace{1cm}
\noindent
{\bf Acknowledgements.} This work was carried out with the support of DFG Grant
KR1401/5-1. The work of JAG was supported by a DFG Mercator Fellowship and he 
thanks the Mathematical Physics Group at Humboldt University, Berlin for 
hospitality. The figures were prepared with {\sc Axodraw} \cite{43}.

\appendix

\sect{Projection matrices.}

In this appendix we record the projection matrices for the computation of
$\Gamma^{g\bar{c}c}_{\mu\nu}(p,q,r)$ and
$\Gamma^{c\bar{c}gg\,abcd}_{\mu\nu\sigma}(p,q,r,s)$ where the formalism was 
introduced earlier. For the former if we define the related projection matrix 
$\tilde{\cal M}^{g\bar{c}c}$ with a common factor removed by
\begin{equation}
{\cal M}^{g\bar{c}c} ~=~ \frac{1}{4[d-2]\Delta_G^2} \tilde{\cal M}^{g\bar{c}c} 
\end{equation}
then the elements are 
\begin{eqnarray}
\tilde{\cal M}^{g\bar{c}c}_{11} &=& 4 [x^2 - 2 x y - 2 x + y^2 - 2 y + 1]^2 
~~,~~ 
\tilde{\cal M}^{g\bar{c}c}_{12} ~=~ -~ 16 [[y - 1]^2 + x^2 - 2 [y + 1] x] y 
\nonumber \\
\tilde{\cal M}^{g\bar{c}c}_{13} &=& \tilde{\cal M}^{g\bar{c}c}_{14} ~=~
-~ 8 [x^2 - 2 x y - 2 x + y^2 - 2 y + 1] [x + y - 1] 
\nonumber \\
\tilde{\cal M}^{g\bar{c}c}_{15} &=& 
-~ 16 [[y - 1]^2 + x^2 - 2 [y + 1] x] x ~~,~~
\tilde{\cal M}^{g\bar{c}c}_{22} ~=~ 64 [d - 1] y^2 
\nonumber \\
\tilde{\cal M}^{g\bar{c}c}_{23} &=& \tilde{\cal M}^{g\bar{c}c}_{24} ~=~
32 [d - 1] [x + y - 1] y 
\nonumber \\
\tilde{\cal M}^{g\bar{c}c}_{25} &=& 
-~ 16 [2 [x^2 - 2 x + y^2 - 2 y + 1] - [x + y - 1]^2 d] 
\nonumber \\
\tilde{\cal M}^{g\bar{c}c}_{33} &=& 
16 [[y - 1]^2 + x^2 + 4 d x y - 2 [3 y + 1] x] 
~~,~~
\tilde{\cal M}^{g\bar{c}c}_{34} ~=~ 16 [d - 1] [x + y - 1]^2 
\nonumber \\
\tilde{\cal M}^{g\bar{c}c}_{35} &=& 32 [y - 1 + x] [d - 1] x ~~,~~
\tilde{\cal M}^{g\bar{c}c}_{44} ~=~ 
16 [[y - 1]^2 + x^2 + 4 d x y - 2 [3 y + 1] x] 
\nonumber \\
\tilde{\cal M}^{g\bar{c}c}_{45} &=& 32 [y - 1 + x] [d - 1] x ~~,~~
\tilde{\cal M}^{g\bar{c}c}_{55} ~=~ 64 [d - 1] x^2 ~. 
\end{eqnarray}
As ${\cal M}^{g\bar{c}c}$ is by construction a symmetric matrix we have only
provided the upper triangle entries. 

For the $4$-point ghost-gluon function there are $36$ Lorentz tensors in the
projection basis for each orientation. They are 
\begin{eqnarray}
{\cal P}^{c\bar{c}gg}_{(1)\mu\nu\sigma} &=& \eta_{\mu\nu} p_\sigma ~~,~~ 
{\cal P}^{c\bar{c}gg}_{(2)\mu\nu\sigma} ~=~ \eta_{\mu\nu} q_\sigma ~~,~~ 
{\cal P}^{c\bar{c}gg}_{(3)\mu\nu\sigma} ~=~ \eta_{\mu\nu} r_\sigma ~~,~~ 
{\cal P}^{c\bar{c}gg}_{(4)\mu\nu\sigma} ~=~ \eta_{\mu\sigma} p_\nu \nonumber \\
{\cal P}^{c\bar{c}gg}_{(5)\mu\nu\sigma} &=& \eta_{\mu\sigma} q_\nu ~~,~~ 
{\cal P}^{c\bar{c}gg}_{(6)\mu\nu\sigma} ~=~ \eta_{\mu\sigma} r_\nu ~~,~~ 
{\cal P}^{c\bar{c}gg}_{(7)\mu\nu\sigma} ~=~ \eta_{\nu\sigma} p_\mu ~~,~~ 
{\cal P}^{c\bar{c}gg}_{(8)\mu\nu\sigma} ~=~ \eta_{\nu\sigma} q_\mu \nonumber \\
{\cal P}^{c\bar{c}gg}_{(9)\mu\nu\sigma} &=& \eta_{\nu\sigma} r_\mu ~~,~~ 
{\cal P}^{c\bar{c}gg}_{(10)\mu\nu\sigma} ~=~ 
\frac{p_\mu p_\nu p_\sigma}{\mu^2} ~~,~~ 
{\cal P}^{c\bar{c}gg}_{(11)\mu\nu\sigma} ~=~ 
\frac{p_\mu p_\nu q_\sigma}{\mu^2} ~~,~~ 
{\cal P}^{c\bar{c}gg}_{(12)\mu\nu\sigma} ~=~ 
\frac{p_\mu p_\nu r_\sigma}{\mu^2}
\nonumber \\
{\cal P}^{c\bar{c}gg}_{(13)\mu\nu\sigma} &=& 
\frac{p_\mu p_\sigma q_\nu}{\mu^2} ~~,~~ 
{\cal P}^{c\bar{c}gg}_{(14)\mu\nu\sigma} ~=~ 
\frac{p_\mu p_\sigma r_\nu}{\mu^2} ~~,~~ 
{\cal P}^{c\bar{c}gg}_{(15)\mu\nu\sigma} ~=~ 
\frac{p_\mu q_\nu q_\sigma}{\mu^2} ~~,~~ 
{\cal P}^{c\bar{c}gg}_{(16)\mu\nu\sigma} ~=~ 
\frac{p_\mu q_\nu r_\sigma}{\mu^2}
\nonumber \\
{\cal P}^{c\bar{c}gg}_{(17)\mu\nu\sigma} &=& 
\frac{p_\mu q_\sigma r_\nu}{\mu^2} ~~,~~ 
{\cal P}^{c\bar{c}gg}_{(18)\mu\nu\sigma} ~=~ 
\frac{p_\mu r_\nu r_\sigma}{\mu^2} ~~,~~ 
{\cal P}^{c\bar{c}gg}_{(19)\mu\nu\sigma} ~=~ 
\frac{p_\nu p_\sigma q_\mu}{\mu^2} ~~,~~ 
{\cal P}^{c\bar{c}gg}_{(20)\mu\nu\sigma} ~=~ 
\frac{p_\nu p_\sigma r_\mu}{\mu^2}
\nonumber \\
{\cal P}^{c\bar{c}gg}_{(21)\mu\nu\sigma} &=& 
\frac{p_\nu q_\mu q_\sigma}{\mu^2} ~~,~~ 
{\cal P}^{c\bar{c}gg}_{(22)\mu\nu\sigma} ~=~ 
\frac{p_\nu q_\mu r_\sigma}{\mu^2} ~~,~~ 
{\cal P}^{c\bar{c}gg}_{(23)\mu\nu\sigma} ~=~ 
\frac{p_\nu q_\sigma r_\mu}{\mu^2} ~~,~~ 
{\cal P}^{c\bar{c}gg}_{(24)\mu\nu\sigma} ~=~ 
\frac{p_\nu r_\mu r_\sigma}{\mu^2}
\nonumber \\
{\cal P}^{c\bar{c}gg}_{(25)\mu\nu\sigma} &=& 
\frac{p_\sigma q_\mu q_\nu}{\mu^2} ~~,~~ 
{\cal P}^{c\bar{c}gg}_{(26)\mu\nu\sigma} ~=~ 
\frac{p_\sigma q_\mu r_\nu}{\mu^2} ~~,~~ 
{\cal P}^{c\bar{c}gg}_{(27)\mu\nu\sigma} ~=~ 
\frac{p_\sigma q_\nu r_\mu}{\mu^2} ~~,~~ 
{\cal P}^{c\bar{c}gg}_{(28)\mu\nu\sigma} ~=~ 
\frac{p_\sigma r_\mu r_\nu}{\mu^2}
\nonumber \\
{\cal P}^{c\bar{c}gg}_{(29)\mu\nu\sigma} &=& 
\frac{q_\mu q_\nu q_\sigma}{\mu^2} ~~,~~ 
{\cal P}^{c\bar{c}gg}_{(30)\mu\nu\sigma} ~=~ 
\frac{q_\mu q_\nu r_\sigma}{\mu^2} ~~,~~ 
{\cal P}^{c\bar{c}gg}_{(31)\mu\nu\sigma} ~=~ 
\frac{q_\mu q_\sigma r_\nu}{\mu^2} ~~,~~ 
{\cal P}^{c\bar{c}gg}_{(32)\mu\nu\sigma} ~=~ 
\frac{q_\mu r_\nu r_\sigma}{\mu^2}
\nonumber \\
{\cal P}^{c\bar{c}gg}_{(33)\mu\nu\sigma} &=& 
\frac{q_\nu q_\sigma r_\mu}{\mu^2} ~~,~~ 
{\cal P}^{c\bar{c}gg}_{(34)\mu\nu\sigma} ~=~ 
\frac{q_\nu r_\mu r_\sigma}{\mu^2} ~~,~~ 
{\cal P}^{c\bar{c}gg}_{(35)\mu\nu\sigma} ~=~ 
\frac{q_\sigma r_\mu r_\nu}{\mu^2}
\nonumber \\
{\cal P}^{c\bar{c}gg}_{(36)\mu\nu\sigma} &=& 
\frac{r_\mu r_\nu r_\sigma}{\mu^2} 
\end{eqnarray}
where we have suppressed the argument. The symmetric projection matrix is 
constructed in the same way as for the $3$-point case. First defining the 
related matrix $\tilde{\cal M}^{c\bar{c}gg}$ by
\begin{equation}
{\cal M}^{c\bar{c}gg} ~=~ \frac{1}{64[d-3]} \tilde{\cal M}^{c\bar{c}gg}
\end{equation}
then the upper triangle elements are 

in $d$-dimensions. Useful in deriving these was the symbolic manipulation
language {\sc Reduce}, \cite{44}. 

\sect{Group theory.}

In this Appendix we summarize aspects of the colour group theory used in 
examining the $4$-point identity. For that example we have concentrated
exclusively on the $SU(\Nc)$ case due to the presence of rank $4$ colour
Casimirs. These arise in $4$-point box graphs through the general fully
symmetric tensors, \cite{36},
\begin{equation}
d_F^{abcd} ~=~ \frac{1}{6} \mbox{Tr} \left( T^a T^{(b} T^c T^{d)} \right) ~~,~~
d_A^{abcd} ~=~ \frac{1}{6} \mbox{Tr} \left( T_A^a T_A^{(b} T_A^c T_A^{d)}
\right)
\label{defrank4}
\end{equation}
where $T^a$ are the group generators and the subscripts $F$ and $A$ indicate 
the fundamental and adjoint representations respectively. These can be related 
to the structure constants $f^{abc}$ and the fully symmetric $SU(\Nc)$ tensor 
$d^{abc}$ via the $SU(\Nc)$ relation
\begin{equation}
T^a T^b ~=~ \frac{1}{2\Nc} \delta^{ab} ~+~ \frac{1}{2} d^{abc} T^c ~+~
\frac{i}{2} f^{abc} T^c ~.
\end{equation}
When there is a contracted tensor product of generators we use
\begin{equation}
T^a_{IJ} T^a_{KL} ~=~ \frac{1}{2} \left[ \delta_{IL} \delta_{KJ} ~-~ 
\frac{1}{\Nc} \delta_{IJ} \delta_{KL} \right] ~.
\end{equation}
In studying $4$-point identity one has more than one colour tensor that can
appear in the Green's functions contributing to (\ref{4ptid}). This is in
contrast to the $3$-point case where $f^{abc}$ is the only structure that
appears a low loop order. Therefore for (\ref{4ptid}) we have to have a basis 
which spans the colour space. If we define the tensors
\begin{equation}
f_4^{abcd} ~\equiv~ f^{abe} f^{cde} ~~~,~~~ 
d_4^{abcd} ~\equiv~ d^{abe} d^{cde} 
\label{ten4def}
\end{equation}
then the second object is not independent of $d_F^{abcd}$ or $d_A^{abcd}$
since for instance
\begin{equation}
f_4^{abcd} ~=~ \frac{2}{\Nc} \left[ \delta^{ac} \delta^{bd} ~-~ \delta^{ad}
\delta^{bc} \right] ~+~ d_4^{acbd} ~-~ d_4^{adbc} ~.
\end{equation}
using results from \cite{11,45}. In \cite{21} the mapping from the non-fully 
symmetric tensor $d_4^{abcd}$ was constructed with for example
\begin{eqnarray}
d_4^{abcd} &=& -~ \frac{1}{3} \left[ f_4^{abcd} - \frac{2}{\Nc}
\left[ \delta^{ac} \delta^{bd} - \delta^{ad} \delta^{bc} \right] \right]
+ \frac{2}{3} \left[ f_4^{acbd} - \frac{2}{\Nc}
\left[ \delta^{ab} \delta^{cd} - \delta^{ad} \delta^{bc} \right] \right]
\nonumber \\
&& +~ \frac{4}{\Nc} \left[ d_A^{acbd} - \frac{2}{3}
\left[ \delta^{ab} \delta^{cd} + \delta^{ac} \delta^{bd}
+ \delta^{ad} \delta^{bc} \right] \right]
\end{eqnarray}
for purely gluonic or ghost boxes and
\begin{eqnarray}
d_4^{abcd} &=& -~ \frac{1}{3} \left[ f_4^{abcd} - \frac{2}{\Nc}
\left[ \delta^{ac} \delta^{bd} - \delta^{ad} \delta^{bc} \right] \right]
+ \frac{2}{3} \left[ f_4^{acbd} - \frac{2}{\Nc}
\left[ \delta^{ab} \delta^{cd} - \delta^{ad} \delta^{bc} \right] \right]
\nonumber \\
&& +~ 8 \left[ d_F^{acbd} - \frac{1}{12\Nc}
\left[ \delta^{ab} \delta^{cd} + \delta^{ac} \delta^{bd} 
+ \delta^{ad} \delta^{bc} \right] \right]
\end{eqnarray}
for boxes involving quarks only. The use of $d_F^{abcd}$ and $d_A^{abcd}$ is
more natural rather than $d_4^{abcd}$ given the fully symmetric nature of the 
$4$-point identity.

\sect{Gluon vertex function examples.} 

The full aribitrary gauge expression of the $3$-gluon vertex function at the 
symmetric point is given by
\begin{eqnarray}
-~ i \left. \Gamma^{ggg}_{\mu\nu\sigma}(p,q,r) \right|_{x=y=1} &=&
\left[ 
\eta_{\mu\nu} q_\sigma 
- \eta_{\mu\nu} p_\sigma 
+ 2 \eta_{\mu\sigma} p_\nu 
+ \eta_{\mu\sigma} q_\nu 
- \eta_{\nu\sigma} p_\mu 
- 2 \eta_{\nu\sigma} q_\mu 
\right] g 
\nonumber \\
&& 
+ \left[
\left[
\left[
- \frac{4}{3} T_F \Nf
+ \frac{2}{3} C_A
+ \frac{3}{4} \xi C_A
\right]
\frac{1}{\epsilon}
- 2 T_F \Nf
+ \frac{4}{3} C_A
- \frac{32}{81} \pi^2 T_F \Nf
\right. \right. \nonumber \\
&& \left. \left. ~~~~~~
+ \frac{1}{81} \pi^2 C_A
- \frac{3}{2} \xi C_A
+ \frac{5}{54} \xi \pi^2 C_A
+ \frac{1}{2} \xi^2 C_A
+ \frac{1}{27} \xi^2 \pi^2 C_A
\right. \right. \nonumber \\
&& \left. \left. ~~~~~~
+ \frac{1}{24} \xi^3 C_A
+ \frac{16}{27} \psi^\prime(\third) T_F \Nf
- \frac{1}{54} \psi^\prime(\third) C_A
- \frac{5}{36} \psi^\prime(\third) \xi C_A
\right. \right. \nonumber \\
&& \left. \left. ~~~~~~
- \frac{1}{18} \psi^\prime(\third) \xi^2 C_A
\right]
\eta_{\mu\nu} p_\sigma
\right. \nonumber \\
&& \left. ~~~~~
+
\left[
\left[
\frac{4}{3} T_F \Nf
- \frac{2}{3} C_A
- \frac{3}{4} \xi C_A
\right]
\frac{1}{\epsilon}
+ 2 T_F \Nf
- \frac{4}{3} C_A
\right. \right. \nonumber \\
&& \left. \left. ~~~~~~~~~
+ \frac{32}{81} \pi^2 T_F \Nf
- \frac{1}{81} \pi^2 C_A
+ \frac{3}{2} \xi C_A
- \frac{5}{54} \xi \pi^2 C_A
- \frac{1}{2} \xi^2 C_A
\right. \right. \nonumber \\
&& \left. \left. ~~~~~~~~~
- \frac{1}{27} \xi^2 \pi^2 C_A
- \frac{1}{24} \xi^3 C_A
- \frac{16}{27} \psi^\prime(\third) T_F \Nf
+ \frac{1}{54} \psi^\prime(\third) C_A
\right. \right. \nonumber \\
&& \left. \left. ~~~~~~~~~
+ \frac{5}{36} \psi^\prime(\third) \xi C_A
+ \frac{1}{18} \psi^\prime(\third) \xi^2 C_A
\right]
\eta_{\mu\nu} q_\sigma
\right. \nonumber \\
&& \left. ~~~~~
+
\left[
\left[
\frac{8}{3} T_F \Nf
- \frac{4}{3} C_A
- \frac{3}{2} \xi C_A
\right]
\frac{1}{\epsilon}
+ 4 T_F \Nf
- \frac{8}{3} C_A
\right. \right. \nonumber \\
&& \left. \left. ~~~~~~~~~
+ \frac{64}{81} \pi^2 T_F \Nf
- \frac{2}{81} \pi^2 C_A
+ 3 \xi C_A
- \frac{5}{27} \xi \pi^2 C_A
- \xi^2 C_A
\right. \right. \nonumber \\
&& \left. \left. ~~~~~~~~~
- \frac{2}{27} \xi^2 \pi^2 C_A
- \frac{1}{12} \xi^3 C_A
- \frac{32}{27} \psi^\prime(\third) T_F \Nf
+ \frac{1}{27} \psi^\prime(\third) C_A
\right. \right. \nonumber \\
&& \left. \left. ~~~~~~~~~
+ \frac{5}{18} \psi^\prime(\third) \xi C_A
+ \frac{1}{9} \psi^\prime(\third) \xi^2 C_A
\right]
\eta_{\mu\sigma} p_\nu
\right. \nonumber \\
&& \left. ~~~~~
+
\left[
\left[
\frac{4}{3} T_F \Nf
- \frac{2}{3} C_A
- \frac{3}{4} \xi C_A
\right]
\frac{1}{\epsilon}
+ 2 T_F \Nf
- \frac{4}{3} C_A
\right. \right. \nonumber \\
&& \left. \left. ~~~~~~~~~
+ \frac{32}{81} \pi^2 T_F \Nf
- \frac{1}{81} \pi^2 C_A
+ \frac{3}{2} \xi C_A
- \frac{5}{54} \xi \pi^2 C_A
- \frac{1}{2} \xi^2 C_A
\right. \right. \nonumber \\
&& \left. \left. ~~~~~~~~~
- \frac{1}{27} \xi^2 \pi^2 C_A
- \frac{1}{24} \xi^3 C_A
- \frac{16}{27} \psi^\prime(\third) T_F \Nf
+ \frac{1}{54} \psi^\prime(\third) C_A
\right. \right. \nonumber \\
&& \left. \left. ~~~~~~~~~
+ \frac{5}{36} \psi^\prime(\third) \xi C_A
+ \frac{1}{18} \psi^\prime(\third) \xi^2 C_A
\right]
\eta_{\mu\sigma} q_\nu
\right. \nonumber \\
&& \left. ~~~~~
+
\left[
\left[
- \frac{4}{3} T_F \Nf
+ \frac{2}{3} C_A
+ \frac{3}{4} \xi C_A
\right]
\frac{1}{\epsilon}
- 2 T_F \Nf
+ \frac{4}{3} C_A
\right. \right. \nonumber \\
&& \left. \left. ~~~~~~~~~
- \frac{32}{81} \pi^2 T_F \Nf
+ \frac{1}{81} \pi^2 C_A
- \frac{3}{2} \xi C_A
+ \frac{5}{54} \xi \pi^2 C_A
+ \frac{1}{2} \xi^2 C_A
\right. \right. \nonumber \\
&& \left. \left. ~~~~~~~~~
+ \frac{1}{27} \xi^2 \pi^2 C_A
+ \frac{1}{24} \xi^3 C_A
+ \frac{16}{27} \psi^\prime(\third) T_F \Nf
- \frac{1}{54} \psi^\prime(\third) C_A
\right. \right. \nonumber \\
&& \left. \left. ~~~~~~~~~
- \frac{5}{36} \psi^\prime(\third) \xi C_A
- \frac{1}{18} \psi^\prime(\third) \xi^2 C_A
\right]
\eta_{\nu\sigma} p_\mu
\right. \nonumber \\
&& \left. ~~~~~
+
\left[
\left[
- \frac{8}{3} T_F \Nf
+ \frac{4}{3} C_A
+ \frac{3}{2} \xi C_A
\right]
\frac{1}{\epsilon}
- 4 T_F \Nf
+ \frac{8}{3} C_A
\right. \right. \nonumber \\
&& \left. \left. ~~~~~~~~~
- \frac{64}{81} \pi^2 T_F \Nf
+ \frac{2}{81} \pi^2 C_A
- 3 \xi C_A
+ \frac{5}{27} \xi \pi^2 C_A
+ \xi^2 C_A
\right. \right. \nonumber \\
&& \left. \left. ~~~~~~~~~
+ \frac{2}{27} \xi^2 \pi^2 C_A
+ \frac{1}{12} \xi^3 C_A
+ \frac{32}{27} \psi^\prime(\third) T_F \Nf
- \frac{1}{27} \psi^\prime(\third) C_A
\right. \right. \nonumber \\
&& \left. \left. ~~~~~~~~~
- \frac{5}{18} \psi^\prime(\third) \xi C_A
- \frac{1}{9} \psi^\prime(\third) \xi^2 C_A
\right]
\eta_{\nu\sigma} q_\mu
\right. \nonumber \\
&& \left. ~~~~~
+ 
\left[
\frac{16}{27} T_F \Nf
- \frac{8}{27} C_A
+ \frac{128}{243} \pi^2 T_F \Nf
- \frac{64}{243} \pi^2 C_A
+ \xi C_A
\right. \right. \nonumber \\
&& \left. \left. ~~~~~~~~~
- \frac{2}{27} \xi \pi^2 C_A
+ \frac{1}{12} \xi^2 C_A
+ \frac{1}{27} \xi^2 \pi^2 C_A
+ \frac{1}{9} \xi^3 C_A
\right. \right. \nonumber \\
&& \left. \left. ~~~~~~~~~
+ \frac{2}{81} \xi^3 \pi^2 C_A
- \frac{64}{81} \psi^\prime(\third) T_F \Nf
+ \frac{32}{81} \psi^\prime(\third) C_A
+ \frac{1}{9} \psi^\prime(\third) \xi C_A
\right. \right. \nonumber \\
&& \left. \left. ~~~~~~~~~
- \frac{1}{18} \psi^\prime(\third) \xi^2 C_A
- \frac{1}{27} \psi^\prime(\third) \xi^3 C_A
\right]
\frac{p_\mu p_\nu p_\sigma}{\mu^2}
\right. \nonumber \\
&& \left. ~~~~~
+ 
\left[
- \frac{28}{27} T_F \Nf
+ \frac{14}{27} C_A
+ \frac{64}{243} \pi^2 T_F \Nf
- \frac{32}{243} \pi^2 C_A
\right. \right. \nonumber \\
&& \left. \left. ~~~~~~~~~
+ \frac{3}{2} \xi C_A
+ \frac{1}{27} \xi \pi^2 C_A
- \frac{7}{12} \xi^2 C_A
- \frac{1}{9} \xi^2 \pi^2 C_A
\right. \right. \nonumber \\
&& \left. \left. ~~~~~~~~~
- \frac{7}{36} \xi^3 C_A
- \frac{2}{81} \xi^3 \pi^2 C_A
- \frac{32}{81} \psi^\prime(\third) T_F \Nf
+ \frac{16}{81} \psi^\prime(\third) C_A
\right. \right. \nonumber \\
&& \left. \left. ~~~~~~~~~
- \frac{1}{18} \psi^\prime(\third) \xi C_A
+ \frac{1}{6} \psi^\prime(\third) \xi^2 C_A
+ \frac{1}{27} \psi^\prime(\third) \xi^3 C_A
\right]
\frac{p_\mu p_\nu q_\sigma}{\mu^2}
\right. \nonumber \\
&& \left. ~~~~~
+ 
\left[
\frac{8}{27} T_F \Nf
- \frac{4}{27} C_A
+ \frac{64}{243} \pi^2 T_F \Nf
- \frac{32}{243} \pi^2 C_A
\right. \right. \nonumber \\
&& \left. \left. ~~~~~~~~~
+ \frac{1}{2} \xi C_A
- \frac{1}{27} \xi \pi^2 C_A
+ \frac{1}{24} \xi^2 C_A
+ \frac{1}{54} \xi^2 \pi^2 C_A
\right. \right. \nonumber \\
&& \left. \left. ~~~~~~~~~
+ \frac{1}{18} \xi^3 C_A
+ \frac{1}{81} \xi^3 \pi^2 C_A
- \frac{32}{81} \psi^\prime(\third) T_F \Nf
+ \frac{16}{81} \psi^\prime(\third) C_A
\right. \right. \nonumber \\
&& \left. \left. ~~~~~~~~~
+ \frac{1}{18} \psi^\prime(\third) \xi C_A
- \frac{1}{36} \psi^\prime(\third) \xi^2 C_A
- \frac{1}{54} \psi^\prime(\third) \xi^3 C_A
\right]
\frac{p_\mu p_\sigma q_\nu}{\mu^2}
\right. \nonumber \\
&& \left. ~~~~~
+ 
\left[
- \frac{8}{27} T_F \Nf
+ \frac{4}{27} C_A
- \frac{64}{243} \pi^2 T_F \Nf
+ \frac{32}{243} \pi^2 C_A
\right. \right. \nonumber \\
&& \left. \left. ~~~~~~~~~
- \frac{1}{2} \xi C_A
+ \frac{1}{27} \xi \pi^2 C_A
- \frac{1}{24} \xi^2 C_A
- \frac{1}{54} \xi^2 \pi^2 C_A
\right. \right. \nonumber \\
&& \left. \left. ~~~~~~~~~
- \frac{1}{18} \xi^3 C_A
- \frac{1}{81} \xi^3 \pi^2 C_A
+ \frac{32}{81} \psi^\prime(\third) T_F \Nf
- \frac{16}{81} \psi^\prime(\third) C_A
\right. \right. \nonumber \\
&& \left. \left. ~~~~~~~~~
- \frac{1}{18} \psi^\prime(\third) \xi C_A
+ \frac{1}{36} \psi^\prime(\third) \xi^2 C_A
+ \frac{1}{54} \psi^\prime(\third) \xi^3 C_A
\right]
\frac{p_\mu q_\nu q_\sigma}{\mu^2}
\right. \nonumber \\
&& \left. ~~~~~
+ 
\left[
\frac{44}{27} T_F \Nf
- \frac{22}{27} C_A
+ \frac{64}{243} \pi^2 T_F \Nf
- \frac{32}{243} \pi^2 C_A
\right. \right. \nonumber \\
&& \left. \left. ~~~~~~~~~
- \frac{1}{2} \xi C_A
- \frac{1}{9} \xi \pi^2 C_A
+ \frac{2}{3} \xi^2 C_A
+ \frac{4}{27} \xi^2 \pi^2 C_A
\right. \right. \nonumber \\
&& \left. \left. ~~~~~~~~~
+ \frac{11}{36} \xi^3 C_A
+ \frac{4}{81} \xi^3 \pi^2 C_A
- \frac{32}{81} \psi^\prime(\third) T_F \Nf
+ \frac{16}{81} \psi^\prime(\third) C_A
\right. \right. \nonumber \\
&& \left. \left. ~~~~~~~~~
+ \frac{1}{6} \psi^\prime(\third) \xi C_A
- \frac{2}{9} \psi^\prime(\third) \xi^2 C_A
- \frac{2}{27} \psi^\prime(\third) \xi^3 C_A
\right]
\frac{p_\nu p_\sigma q_\mu}{\mu^2}
\right. \nonumber \\
&& \left. ~~~~~
+ 
\left[
- \frac{44}{27} T_F \Nf
+ \frac{22}{27} C_A
- \frac{64}{243} \pi^2 T_F \Nf
+ \frac{32}{243} \pi^2 C_A
\right. \right. \nonumber \\
&& \left. \left. ~~~~~~~~~
+ \frac{1}{2} \xi C_A
+ \frac{1}{9} \xi \pi^2 C_A
- \frac{2}{3} \xi^2 C_A
- \frac{4}{27} \xi^2 \pi^2 C_A
\right. \right. \nonumber \\
&& \left. \left. ~~~~~~~~~
- \frac{11}{36} \xi^3 C_A
- \frac{4}{81} \xi^3 \pi^2 C_A
+ \frac{32}{81} \psi^\prime(\third) T_F \Nf
- \frac{16}{81} \psi^\prime(\third) C_A
\right. \right. \nonumber \\
&& \left. \left. ~~~~~~~~~
- \frac{1}{6} \psi^\prime(\third) \xi C_A
+ \frac{2}{9} \psi^\prime(\third) \xi^2 C_A
+ \frac{2}{27} \psi^\prime(\third) \xi^3 C_A
\right]
\frac{p_\nu q_\mu q_\sigma}{\mu^2}
\right. \nonumber \\
&& \left. ~~~~~
+ 
\left[
\frac{28}{27} T_F \Nf
- \frac{14}{27} C_A
- \frac{64}{243} \pi^2 T_F \Nf
+ \frac{32}{243} \pi^2 C_A
\right. \right. \nonumber \\
&& \left. \left. ~~~~~~~~~
- \frac{3}{2} \xi C_A
- \frac{1}{27} \xi \pi^2 C_A
+ \frac{7}{12} \xi^2 C_A
+ \frac{1}{9} \xi^2 \pi^2 C_A
\right. \right. \nonumber \\
&& \left. \left. ~~~~~~~~~
+ \frac{7}{36} \xi^3 C_A
+ \frac{2}{81} \xi^3 \pi^2 C_A
+ \frac{32}{81} \psi^\prime(\third) T_F \Nf
- \frac{16}{81} \psi^\prime(\third) C_A
\right. \right. \nonumber \\
&& \left. \left. ~~~~~~~~~
+ \frac{1}{18} \psi^\prime(\third) \xi C_A
- \frac{1}{6} \psi^\prime(\third) \xi^2 C_A
- \frac{1}{27} \psi^\prime(\third) \xi^3 C_A
\right]
\frac{p_\sigma q_\mu q_\nu}{\mu^2}
\right. \nonumber \\
&& \left. ~~~~~
+ 
\left[
- \frac{16}{27} T_F \Nf
+ \frac{8}{27} C_A
- \frac{128}{243} \pi^2 T_F \Nf
+ \frac{64}{243} \pi^2 C_A
\right. \right. \nonumber \\
&& \left. \left. ~~~~~~~~~
- \xi C_A
+ \frac{2}{27} \xi \pi^2 C_A
- \frac{1}{12} \xi^2 C_A
- \frac{1}{27} \xi^2 \pi^2 C_A
\right. \right. \nonumber \\
&& \left. \left. ~~~~~~~~~
- \frac{1}{9} \xi^3 C_A
- \frac{2}{81} \xi^3 \pi^2 C_A
+ \frac{64}{81} \psi^\prime(\third) T_F \Nf
- \frac{32}{81} \psi^\prime(\third) C_A
\right. \right. \nonumber \\
&& \left. \left. ~~~~~~~~~
- \frac{1}{9} \psi^\prime(\third) \xi C_A
+ \frac{1}{18} \psi^\prime(\third) \xi^2 C_A
+ \frac{1}{27} \psi^\prime(\third) \xi^3 C_A
\right]
\frac{q_\mu q_\nu q_\sigma}{\mu^2}
\right] g^3 \nonumber \\
&& +~ O(g^5) 
\end{eqnarray}
in terms of the full tensor basis where we note again that we have used the 
compact notation $\xi$~$=$~$1$~$-$~$\alpha$. Also for assistance at the 
$3$-point symmetric point $x$~$=$~$y$~$=$~$1$ we have 
\begin{equation}
\Phi_1(1,1) ~=~ \frac{4\pi^2}{9} ~-~ \frac{2}{3} \psi^\prime (\third) 
\end{equation}
which is related to the Clausen function since $\Delta_G(1,1)$~$=$~$-$~$3$
leading to a complex value for $\rho(1,1)$ and an imaginary one for
$\lambda(1,1)$. 

For the $4$-point identity we record the contraction of the symmetric point
quartic gluon Green's function with one external momentum to illustrate several
subtle points. The Landau gauge expression restricted to $SU(\Nc)$ is

where we have set 
\begin{equation}
\tilde{\Phi}_1(\threequarters) ~=~ 
\Phi_1 \left( \threequarters, \threequarters \right) ~~~,~~~
\tilde{\Phi}_1(\ninesixteenths) ~=~ 
\Phi_1 \left( \ninesixteenths, \ninesixteenths \right) 
\end{equation}
for shorthand. In \cite{21} the non-contracted expression included the quartic 
colour group Casimir in the fundamental representation $d_F^{abcd}$ of 
(\ref{defrank4}) in addition to the adjoint one. When the full expression for
$\Gamma^{gggg\,abcd}_{\mu\nu\sigma\rho}(p,q,r,s)$, which contains $d_F^{abcd}$,
is contracted with $s^\rho$ to produce (\ref{4ptcon}) it transpires that all 
the $d_F^{abcd}$ terms cancel. This is not unexpected since there are no other 
places in the identity (\ref{4ptid}) for such a Casimir to arise at one loop. 
The places where $d_F^{abcd}$ can potentially appear are in the various 
orientations of $\Gamma^{c\bar{c}gg\,abcd}_{\nu\sigma\lambda}(p,q,r,s)$ and at 
one loop there are no box graphs involving quarks. However 
$\Gamma^{gggg\,abcd}_{\mu\nu\sigma\rho}(p,q,r,s)$ does depend on $\Nf$ through 
the reorganization of the group theory associated with the purely quark boxes.  

While the expressions for each orientation of 
$\Gamma^{c\bar{c}gg\,abcd}_{\mu\nu\sigma}(p,q,r,s)$ is similar we provide that 
for the case of $A$ for purposes of the discussion. In the Landau gauge we have

also for $SU(\Nc)$. Unlike 
$\Gamma^{gggg\,abcd}_{\mu\nu\sigma\rho}(p,q,r,s)$ there are no quark
contributions.

\end{document}